\documentclass[a4paper]{aa}

\usepackage{natbib}
\usepackage{txfonts}
\usepackage{epsfig}

\begin{document} 

\newcommand{\es}{erg s$^{-1}$}   
\newcommand{\ecms}{erg~cm$^{-2}$~s$^{-1}$}
\newcommand{\halpha}{H$\alpha$}  
\newcommand{\hbeta}{H$\beta$}
\newcommand{\kms}{km~s$^{-1}$}   
\newcommand{\cmthree}{cm$^{-3}$}
\newcommand{\msun}{M$_{\odot}$} 
\newcommand{\xmm}{XMM-\emph{Newton}} 
\newcommand{\nh}{\mbox{$N({\rm H})$}}
\newcommand{\chandra}{\emph{Chandra}}

\title{X-ray and optical bursts and flares in YSOs: results from a
5-day \xmm\ monitoring campaign of L1551}

\author{G.\ Giardino \inst{1} \and F.\ Favata\inst{1} \and B.\
Silva\inst{1,3} \and G.\ Micela \inst{2} \and F.\ Reale\inst{4} \and
S.\ Sciortino\inst{2}}

\institute{Astrophysics Division -- Research and Science Support
 Department of ESA, ESTEC, 
  Postbus 299, NL-2200 AG Noordwijk, The Netherlands
\and
  INAF --Osservatorio Astronomico di Palermo, 
 Piazza del Parlamento 1, I-90134 Palermo, Italy 
\and
 Centro de Astrofisica da Universidade do Porto, Rua das Estrelas,
4150-762 Porto, Portugal
\and
Dipartimento di Scienze Fisiche \&
  Astronomiche, Sezione di Astronomia, Piazza del Parlamento 1, 90134
  Palermo Italy
}

\offprints{G. Giardino,\\ Giovanna.Giardino@rssd.esa.int}

\date{Received date / Accepted date}

\titlerunning{X-ray and optical bursts and flares in YSOs}
\authorrunning{}

\abstract{We present the results of a five-day monitoring campaign
  with \xmm\ of six X-ray bright young stellar objects (YSOs) in
  the star-forming complex L1551 in Taurus. All stars present
  significant variability on the five-day time scale. Modulation of
  the light curve on time scales comparable with the star's rotational
  period appeared to be present in the case of one weak-lined T Tauri
  star. Significant spectral variations between the 2000 and the
    2004 observations were detected in the (unresolved) classical T
    Tauri binary system XZ Tau: a hot plasma component which was
    present in the X-ray spectrum in 2000 had significantly weakened
    in 2004. As XZ~Tau N was undergoing a strong optical outburst in
  2000, which had terminated since then, we speculate on the possible
  relationship between episodic, burst accretion, and X-ray heating.
  The transition object HL~Tau underwent a strong flare with a complex
  temperature evolution, which is indicative of an event confined within a very
  large magnetic structure (few stellar radii), similar to the ones
  found in YSOs in the Orion Nebula Cluster.   }

\maketitle

\section{Introduction}
\label{sec:intro}

Young stars are copious sources of X-ray emission; this, coupled with
the fact that X-rays are not strongly affected by dust, makes X-ray
observations a very useful tool in the study of the early phases of
stellar evolution. In particular, X-ray observations of young
stars provide insights into high energy processes, such as coronal
emission, and trace the stars' magnetic activity.  While Class 0
objects have not yet been conclusively detected in the X-ray, X-ray
emission from Class I objects is relatively common; Class II, or
classical T Tauri stars (CTTS) and Class III, or weak-lined T Tauri
stars (WTTS), are almost always intense X-ray sources (e.g.\ 
\citealp{pf2005}).

While X-ray emission is interpreted in WTTS to be coronal in origin (a
scaled-up version of the solar activity with similar energy production
and emission mechanism), the phenomenology for CTTS and Class I
objects is more complex than a simple manifestation of enhanced
solar-type magnetic phenomena, which suggests that accretion and the
presence of relatively massive accretion disks may also play a role.
Evidence of the role of accretion is provided by TW Hya, one of the
nearest CTTS and thus one for which high-resolution X-ray spectra are
currently obtainable. Its inferred plasma temperature distribution,
density, and peculiar chemical abundances are consistent with a model
in which the bulk of the X-ray emission is generated via mass
accretion (\citealp{khs+2002}; \citealp{ss2004} and see also
\citealp{dra2005}). The presence of an accretion funnel shock at the
site of the X-ray and UV emission is also invoked to explain the low
flux in the forbidden line in the O\,{\sc vii} triplet of the \xmm\
RGS spectrum of BP Tau (\citealp{srn+2005}). Nevertheless, the
presence of hot plasma, hotter than possibly generated by the
accretion shock, shows that different X-ray emission processes must be
going on at the same time in accreting YSOs. Statistically,
high-accretion objects have lower $L_{\rm X}$ than non-accreting
ones (e.g.\ in the Orion Nebula Cluster, ONC, \citealp{fdm+2003},
\citealp{pkf+2005}).

Temporal variability is a useful tool for distinguishing between the
underlying X-ray emission processes in different young stellar
types. The observation, for example, of rotationally induced
modulation of X-ray emission probes the spatial distribution of the
emitting plasma and the time scales of its temporal evolution. Until
recently, there were only a handful of reports of rotational
modulation of X-ray emission from young main sequence stars: VXR45, a
young fast rotator star ($P=0.22$~d) member of IC 2391
(\citealp{mmp+2003}); AB Dor, another fast rotator ZAMS K0 dwarf
($P=0.51$~d) (\citealp{hbd+2005}); and EK~Dra (a young sun analogue)
with a period $P=2.7$~d, for which indications of X-ray rotational
modulation are given by \citet{gsb+95}. More recently, rotational
modulation in the X-ray light curve of several pre-main sequence stars
(PMS) (\citealp{fms+2005b}) has been reported, based on the two-week
\chandra\ monitoring campaign of the ONC known as COUP
(\citealp{gfb+2005}).
 

In a previous 50 ks \xmm\ observation of the star-forming complex
L1551 performed in 2000, Favata et al. (2003; hereafter FGM03) report
interesting and unexpected variability in the CTTS
XZ~Tau, apparently neither due to `classical' flaring nor to simple
rotational modulation. To fully understand the observed variability
(whose timescales were clearly under-sampled by the 2000
observation) and the underlying processes, we  performed a
monitoring campaign of L1551 with \xmm, composed of 11 exposures of
roughly 9 ks each regularly spaced over 5 days, for a total
integration time of about 100 ks. Although much more limited in
temporal coverage and total sensitivity than COUP, the L1551 campaign
discussed here probed similar times scales, and the larger collecting
area of \xmm\ allows spectral variations to also be  monitored on
shorter time scales.  Unfortunately the observations were affected (as
discussed in detail later) by rather unfavorable background conditions,
which made it impossible to fulfill all the original goals of the
project.  Nevertheless, we exploited the data to extract the
maximum amount of information possible on the variability of the seven
bright X-ray sources in the field.

The paper is organized as follows: the observations and the data
reduction procedure are presented in Sect.~\ref{sec:obs}, results for
the individual sources are presented in Sect.~\ref{sec:res}, while a
global analysis of their variability is presented in
Sect.~\ref{sec:vari}. The results are discussed in
Sect.~\ref{sec:disc}, while the conclusions are summarized in
Sect.~\ref{sec:concl}.

\section{Observations}
\label{sec:obs}

The \xmm\ observation discussed in this paper consists of 11 exposures
of roughly 9 ks each of the L1551 star-forming cloud. All observations
were pointed in the same direction with the same roll angle, with
bore-sight coordinates of RA 04:31:39, Dec 18:10:00. The exposures
were regularly spaced over 5 days, starting March 4, 2004 at 16:02
UTC.  All three EPIC cameras were active during the observations, in
full-frame mode with the medium filters. The \xmm\ EPIC cameras have a
field of view of approximately 30 arcmin, with a PSF (in the center of
the field of view) with a resolution of approximately 5 arcsec, and a
useful bandpass of 0.3 to 10 keV. The \xmm\ mission is described e.g.\
in \citet{jla+2001}.

Table~\ref{tab:obs_id_details} summarizes  the start time, time length, and livetime, in each camera, for each of the 11 exposure. For MOS1 and
MOS2, the exposure's typical length (elapsed time) is approximately 8
ks, except for obs.\ 0501 and 0901, lasting 10.5 and 13.7 ks,
respectively.  The PN exposures are usually shorter, lasting typically
6--7 ks, with a number of exceptions: obs.\ 0701 and 0801 lasted only
2.5 and 3.4 ks, respectively, and obs. 0501 and 0901 lasted 10.2 and
14.0 ks, respectively. PN is alive for a significantly smaller
fraction of the exposures' duration than MOS1 and MOS2. Over the 11
exposures, the total observation time for PN is 82 ks, while for
MOS1/MOS2, it is 107 ks, with corresponding livetimes of 68 ks and 105 ks.

\begin{figure}[!tbp]
  \begin{center} \leavevmode \epsfig{file=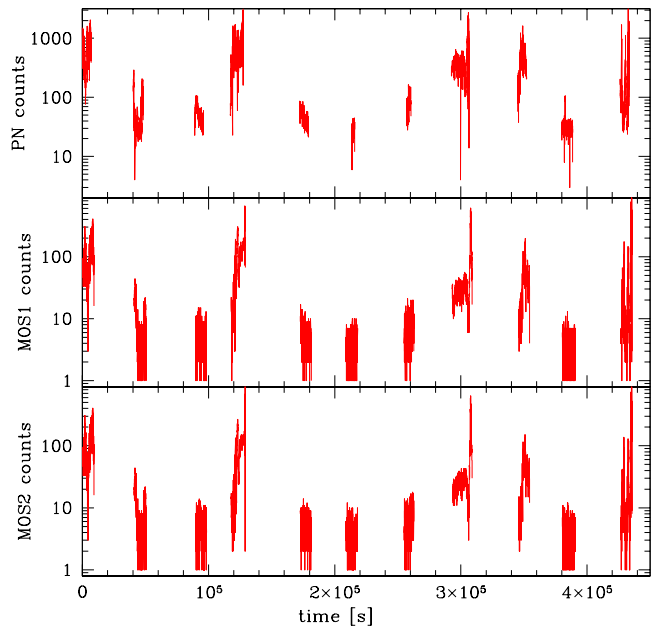, width=9.0cm}
    \caption{The time evolution of the background (total counts from
    the camera at energies $E \ge 8$~keV per 30 s bin) for the entire
    observation, separately plotted for each of the three detectors.
    } \label{fig:backg} \end{center}
\end{figure}

Figure~\ref{fig:backg} shows the light curves of the background (total
counts from each camera at energies above 8 keV) in the three
instruments. Many of the 11 exposures were significantly contaminated
by proton flares, in particular obs.\ 0201, 0501 0901, 1001, and 1201
display a number of intense short duration flares and long duration
(time scales comparable to the exposure times) episodes of high
background. PN data are generally more contaminated than MOS1 and
MOS2. The background level appears to modulated on time scales similar
to the 2 day duration of \xmm\ orbit. The background is high, as shown
by comparison with the 2000 observation, for which only time intervals
with less than 15 background counts per 30 sec bin were
retained. Applying such a strategy to this set of observations would
result in the rejection of the whole 5-day campaign. The cleanest
segments here have background levels that are at least twice as high,
and the most contaminated ones reach background count rates up to 30
times higher.

To recover the maximum amount of information possible, in spite of the
high background, data were processed in two different ways, one aimed
at determining time-resolved spectral parameters for the X-ray bright
sources, the other aimed at recovering information for the weaker
sources. The latter involved stacking, or merging, the 11 individual
exposures into one single photon list. In both cases, only photons with
energy ranging between 0.3 and 8 keV were retained.

For the spectral analysis of the brighter sources, only time intervals
affected by the strongest background spikes were removed, ensuring
that only a small fraction of the three detectors' livetime was
discarded from each exposure (never more than 10\%). This implies that
the background level in individual exposures will be very different;
while not fully satisfactory, this is the only strategy that allows us to
maintain the time coverage of the original observation. The
effectiveness of background subtraction has been tested by verifying
the consistency of the derived spectral parameters for different
levels of background filtering. 

The data were processed with the standard SAS V6.0.0 pipeline.
Source and background photons were extracted from the filtered event
files using a set of scripts developed at Palermo
Observatory. Source and background regions were defined interactively
in each exposure separately using the \textsc{ds9} display software.
The background was extracted from regions on the same CCD chip and at
a similar off-axis angle to the source region.  Response matrices
(``\textsc{rmf} and \textsc{arf} files'') appropriate to the position
and size of the source extraction regions were computed.

The spectral analysis was performed using the \textsc{xspec}
package V11.2, after rebinning the source spectra to a minimum of 20
source counts per (variable width) spectral bin.  The spectral fits
were all carried out in the energy range 0.3--7.5~keV, unless
otherwise stated.  All light curves and spectra discussed in this
paper were background subtracted.  Since V827 Tau and V1075 Tau were
unavailable in the two MOS cameras\footnote{V827~Tau fell out of MOS1
  and was close to the camera edge in MOS2, V1075 fell out of MOS1
  and, because of its weakness, it is hardly visible in some MOS2
  exposures.} and the MOS1 and MOS2 data in obs.\ 0901 were corrupted
and unusable, all light curves presented in the following section are
from PN data. For the spectral analysis, we used PN data for all the
sources except for V710~Tau, for which we used the results from
simultaneous fits to PN, MOS1, and MOS2 spectral data (for all
exposures apart 0901) due to the source's weakness. For the 5 sources for
which data from all three cameras are available, joint fits to PN,
MOS1, and MOS2 spectral data yielded spectral parameters compatible
with the ones derived from PN data alone.

To properly compare the present observation with the 2000 one, we also
reprocessed the 2000 data with the new SAS pipeline V6.0.0. This
resulted in a number of somewhat disturbing differences; in particular,
for two sources (V827 Tau and V710 Tau, discussed in detail in
Sect.~\ref{sec:res}), for which apparently valid spectra resulted from
the V5.3.1 pipeline used by FGM03, no `good' photons are left in the
results of the V6.0.0 pipeline. Indeed, the fluxes derived by FGM03
for V827 Tau and V710 Tau are very different from the ones derived
here for the 2004 observation, a discrepancy evidently caused by the
problems in the V5.3.1 processing. For the other sources, the flux and
spectral parameters derived in FGM03 and in the present work are
generally in good agreement, although some important differences will
be discussed later.

\begin{table*}[!tbh]
  \begin{center}
    \caption{Start time,
      exposure duration and livetime for the 11 individual exposures
      comprising the \xmm\ campaign on L1551.} 

    {\footnotesize
      \begin{tabular}{c|cccc|ccc|ccc}
        \multicolumn{1}{c|}{Obs. ID} & \multicolumn{4}{c|}{Start time
        [yyyy-mm-dd hh-mm-ss UTC]} & \multicolumn{3}{c|}{Elapsed time
        [ks]} & \multicolumn{3}{c}{Livetime [ks]}\\ 
      & & \multicolumn{1}{c}{PN} &\multicolumn{1}{c}{MOS1} &
        \multicolumn{1}{c|}{MOS2} & \multicolumn{1}{c}{PN} &
        \multicolumn{1}{c}{MOS1} & \multicolumn{1}{c|}{MOS2} &  
        \multicolumn{1}{c}{PN} & \multicolumn{1}{c}{MOS1} &
        \multicolumn{1}{c}{MOS2}\\ 
        \hline
      
0201 & 2004-03-04 & 16:25:28 & 16:03:09 & 16:03:08 & 7.0  &  8.7  &  8.7  &  5.9 &  8.6 &  8.6\\
0301 & 2004-03-05 & 03:40:49 & 03:18:32 & 03:18:29 & 7.9  &  9.6  &  9.6  &  7.1 &  9.4 &  9.4\\
0401 & 2004-03-05 & 17:12:35 & 16:50:24 & 16:50:15 & 7.0  &  8.7  &  8.7  &  6.3 &  8.5 &  8.5\\
0501 & 2004-03-06 & 01:03:13 & 00:40:57 & 00:40:52 & 10.2 & 11.2  & 11.2  &  6.9 & 10.9 & 10.9\\
0601 & 2004-03-06 & 16:17:29 & 15:55:10 & 15:55:08 & 7.0  &  8.7  &  8.7  &  6.3 &  8.6 &  8.6\\
0701 & 2004-03-07 & 03:44:21 & 01:56:54 & 01:56:54 & 2.5  &  9.3  &  9.3  &  2.2 &  9.0 &  9.0\\
0801 & 2004-03-07 & 15:53:17 & 14:46:39 & 14:46:39 & 3.4  &  7.8  &  7.7  &  3.0 &  7.4 &  7.4\\
0901 & 2004-03-08 & 01:44:52 & 01:22:32 & 01:22:32 & 14.0 & 15.6  & 15.7  & 10.5 & 15.2 & 15.4\\
1001 & 2004-03-08 & 16:19:29 & 15:57:11 & 15:57:09 & 7.0  &  8.7  &  8.7  &  6.3 &  8.6 &  8.6\\
1101 & 2004-03-09 & 01:59:08 & 01:36:55 & 01:36:49 & 8.7  & 10.4  & 10.4  &  7.7 & 10.1 & 10.1\\
1201 & 2004-03-09 & 14:53:25 & 14:31:13 & 14:31:05 & 7.0  &  8.7  & 8.7  &  6.1 &  8.5 &  8.5\\
      
\end{tabular} }
\label{tab:obs_id_details}
\end{center}
\end{table*}

\subsection{The merged data}

\begin{figure*}[!tbp]
  \begin{center} \leavevmode \epsfig{file=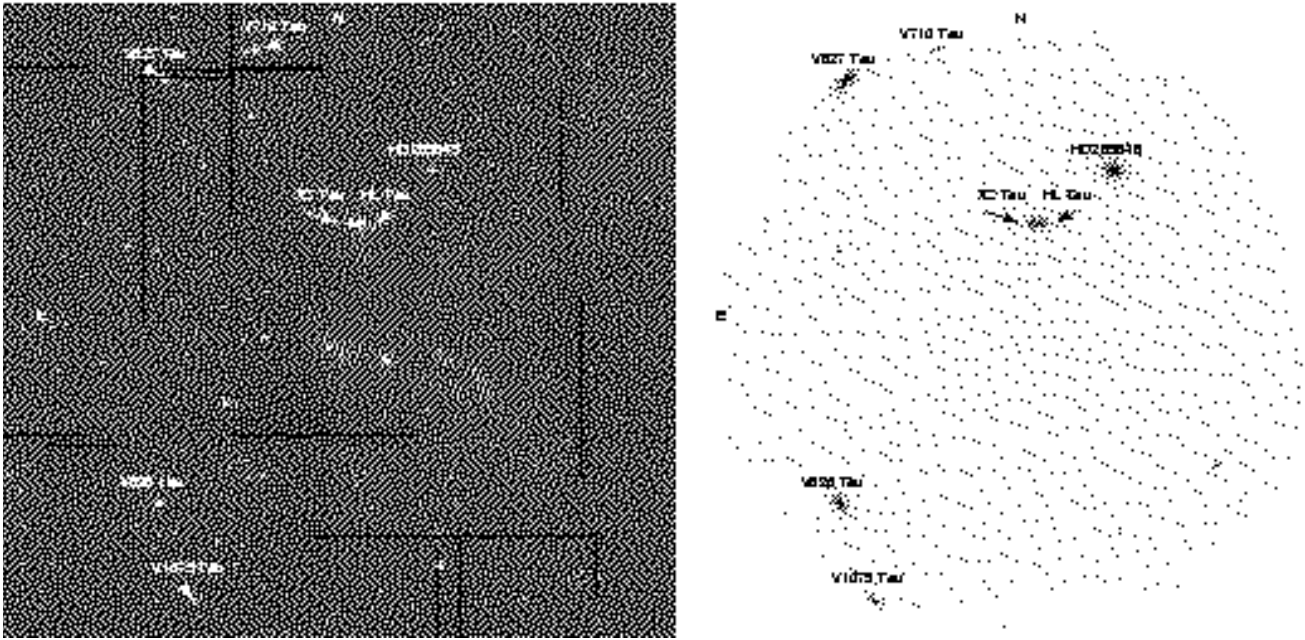, width=18.0cm,
    bbllx=38, bblly=275, bburx=576, bbury=538, clip=}

    \caption{The left panel shows a mosaic of 2MASS $K$-band images of
    the region observed by \xmm, while the right panel shows an X-ray
    image produced with the merged low-background data from all three
    EPIC detectors. The two images are on the same scale and have been
    registered (field size $\sim 30' \times 30'$ centered at 04:31:44
    $+$18:10:29). The sources discussed in the paper have been
    labeled.}
	
	\label{fig:image} 
	\end{center}
\end{figure*}

Most of the known X-ray sources in L1551 are too weak to be visible in
any of the eleven individual exposures. To try to detect fainter
sources in the field, we stacked the individual exposures for each
EPIC camera using the task {\sc merge} in the SAS V6.0.0
pipeline. Prior to stacking, each exposure was filtered to reduce the
background. To recover faint sources, the background must be reduced
significantly below the high values present in the data; the 11
exposures were filtered (before merging) at a threshold level of 100
background counts per 30 sec bin for PN and 60 counts per 30 sec bin
for MOS1 and MOS2. Table \ref{tab:merge_times} summarizes the total
(merged) live time before and after filtering. The PN camera was the
one most affected by proton flares, and the restrictive filtering
applied discards nearly half of the data. Thus, the remaining 'clean'
(low background) data in this observation is only $2/3$ for PN and
just over $4/3$ for MOS1 and MOS2 of the clean time in the observation
analyzed by FGM03 (even though the total exposure is twice as
long). The observation is therefore not deeper than the one of FGM03.
We used the merged observation to derive integrated spectra of 5 of
the 7 bright sources (the ones that did not undergo a flare during the
observation) and compared the resulting spectral parameters with the
average value of the spectral parameters derived from the individual
exposures. The good agreement between the two is an important
verification of the validity of the spectra derived from the single
exposures, despite the high background levels. An X-ray image produced
from the merged data is shown in Fig.~\ref{fig:image}, together with a
$K$-band 2MASS image.

\begin{table}[h]

\caption{Filtering thresholds (background counts per 30 sec bin) and
livetime before and after filtering for the merged data for the three
detectors. See text for more details.}

     \begin{flushleft} {\footnotesize 
     \begin{tabular}{c|c|cc}
     \multicolumn{1}{c|}{Merged observations} & \multicolumn{1}{c|}{Filtering [Counts]} & \multicolumn{2}{c}{Livetime [s]} \\
     \multicolumn{1}{c|}{Instrument} & $E > 8 $~keV & \multicolumn{1}{c}{Before} &\multicolumn{1}{c}{After}  \\\hline
     
     PN & 100 & 68482 & 35053 \\
     MOS1 & 60 & 89521 & 73483 \\
     MOS2 & 60 & 89660 & 74994 \\
    
     \hline
     \end{tabular} }
     \end{flushleft}
     \label{tab:merge_times}
     \end{table}

\section{Analysis of the individual sources}
\label{sec:res}

\subsection{V826 Tau}

V826 Tau is a K7 SB2 WTTS (\citealp{mwf+83}). The separation between
the two components is 0.06 AU (\citealp{jmf94}) and the period is 3.9
d (\citealp{mat94}). A photometric period of 3.7 d is known
(\citealp{bck+95}).

\begin{figure}[!tbp]
  \begin{center} \leavevmode \epsfig{file=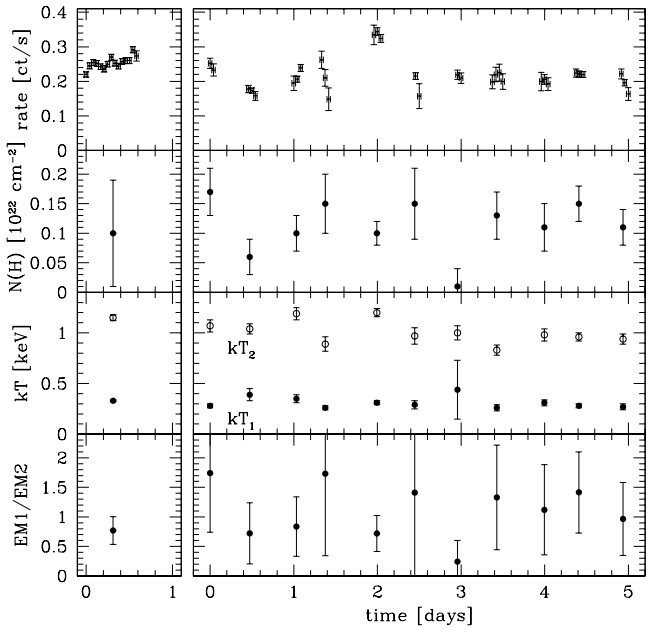, width=9.0cm}
    \caption{{\em Right panel -} From top to bottom: light curve of
      V826 Tau over the 5 days of monitoring (bin-time 3600 s),
      best-fit values for \nh\, $kT_1$, $kT_2$ and for the ratio
      $E\!M_1/E\!M_2$. {\em Left panel -} light curve and best-fit
      values of spectral parameters for V826~Tau from the \xmm\ 
      observation of Sept. 2000 (from FGM03). All data are from the PN
      camera.}
    \label{fig:v826all}
  \end{center}
\end{figure}

During the 2000 \xmm\ observations (FGM03), the count rate of V826~Tau
increased by $\simeq 50$\% over the 50 ksec. Also during this 
5-day monitoring campaign, V826~Tau showed significant
variability on different time scales (Fig.~\ref{fig:v826all}) with a
factor of $\simeq 2$ amplitude.

The spectral parameters for each exposure are summarized in
Table~\ref{tab:v826_all11pn}. All PN spectra could be fit with
absorbed 2T plasma models with metallicity frozen at 0.17~$Z_{\odot}$
(the best-fit value obtained by FGM03). The count rate variability is
not linked to large variations in the spectral parameters. The average
values of the spectral parameters and of the X-ray luminosity are
compatible with the values determined by FGM03.
 
\subsection{V827 Tau}

\begin{figure}[!tbp]
  \begin{center} \leavevmode \epsfig{file=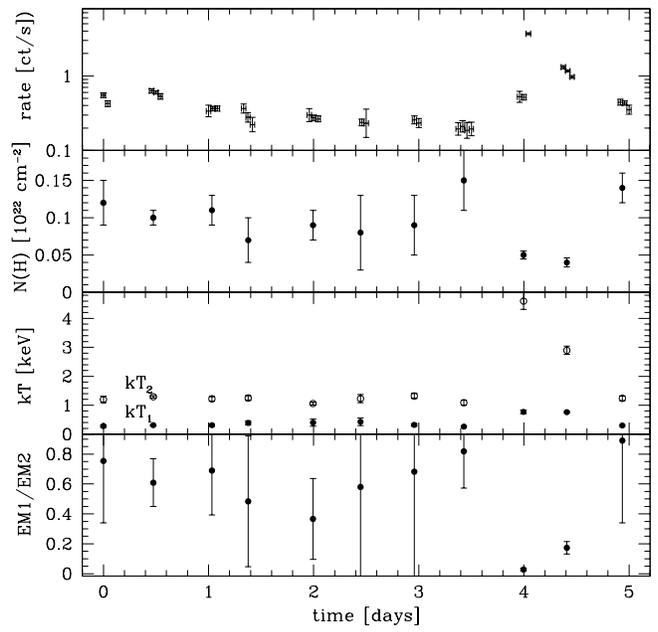, width=9.0cm}
    \caption{From top to bottom: light curve of V827 Tau over the 5
    days of monitoring (bin-time 3600 s) and best-fit values of \nh\,
    $kT_1$, $kT_2$, and $E\!M_1/E\!M_2$ ratio (PN data), for the nine
    exposures not affected by the star's flare. During the star's
    flare the parameter values of the absorbed 3T-plasma fit are
    given.  The PN data from \xmm\ observation of 2000 are not given for
    this source because they were found unreliable after reprocessing
    with more recent SAS pipeline (see text for more details).}
    \label{fig:v827all} \end{center}
\end{figure}

V827 Tau is a K7 X-ray bright WTTS with a rotational period of 3.75 d
(\citealp{bck+95}).  In the light curve (Fig.~\ref{fig:v827all}) a
large flare lasting over one day is present with a count rate
increase of a factor of 10. The emission before the flare varies
significantly, by a factor of 3 over $\simeq$ 3 d.

The best-fit spectral parameters are reported in
Table~\ref{tab:v827_all11pn}. Excluding the flaring intervals, no
significant spectral variations were present. The typical values (\nh\
$\sim 0.1\times 10^{22}~{\rm cm^{-2}}$, $kT_1\sim 0.3$~keV, $kT_2\sim
1.2$~keV, and $Z=0.14~Z_{\odot}$) are very similar to the values
derived in FGM03.

As in Table~\ref{tab:v827_all11pn}, the source intrinsic flux prior
to the flare decreases slowly by a factor of $3$. The minimum X-ray
luminosity is 4 times higher than the value reported by FGM03; however
(as discussed in Sect.~\ref{sec:obs}), reprocessing with the V6.0.0
software has shown that the data for this source were not properly
reduced by the previous version of the SAS pipeline (probably because
the source was very close to the edge of one of the chips); thus the
X-ray luminosity value of V827~Tau given in FGM03 is most likely incorrect.

\subsubsection{Flare analysis}

To determine the flaring emission spectral parameters we 
subdivided the rise phase of the flare (Obs. 1001) into two segments
(one for the rise phase and one for the peak). In order to separate
the quiescent contribution from the flaring emission, we modeled
the spectra with an absorbed 3T-plasma. The parameters of the first
two components (representing the quiescent emission) were frozen to
the values derived by simultaneously fitting four spectra during the
quiescent phase (Obs.\ 0201, 0401, 0601, 0701, 0801). The values for
the third component (the flaring emission) were then fitted. The
results for this component are summarized in
Table~\ref{tab:v827_flare}.

To derive the flare's physical parameters we used the approach
initially discussed by \citet{rbp+97} and since then applied to a
variety of stellar flares. The calibration of the method for the \xmm\
detectors, and a detailed explanation of the physics behind it can be
found in \citet{rgp+2004}, to which the reader is referred. This
approach allowed us to account properly for the presence of sustained
heating during the flare decay, using the slope $\zeta$ of the flare
decay in the $\log T$ vs. $\log \sqrt{E\!M}$ diagram. The semi-length
of the flaring loop in this formulation is given by
\begin{equation}
L = \frac{\tau_{\rm LC} \sqrt{T_{\rm max}}}{\alpha F(\zeta)}~~~~~~~0.35 <
\zeta \le 1.6 
\label{eq:loop}
\end{equation}
where $\alpha = 3.7 \times 10^{-4} {\rm cm^{-1} s^{-1} K^{1/2}}$,
$\tau_{\rm LC}$ is the $1/e$ folding time of the light curve decay,
and $T_{\rm max}$ is the peak temperature of the plasma in the flaring
loop. Then, $F(\zeta)$ and the relationship between $T_{\rm max}$ and the
best-fit peak temperature $T_{\rm obs}$ are both functions that
need to be separately determined for each X-ray detector, depending
on its spectral response. For the EPIC PN,
\begin{equation}
F(\zeta) = c_a/(\zeta - \zeta_a) + q_a
\label{eq:fchi}
\end{equation}
where $c_a =0.51 \pm 0.03$, $\zeta_a = 0.35 \pm 0.01$, and $q_a = 1.36
\pm 0.18$. The range of validity corresponds to an impulsively heated
flare ($\zeta = 1.66$) and to very slow decays (strong sustained
heating) corresponding to the locus of statics loops ($\zeta \le
0.35$). The maximum temperature $T_{\rm max}$ is derived from $T_{\rm
  obs}$ as
\begin{equation}
T_{\rm max} = 0.130 T_{\rm obs}^{1.16}~~~~.
\end{equation}

\begin{figure}[!tbp]
        \begin{center} \leavevmode \epsfig{file=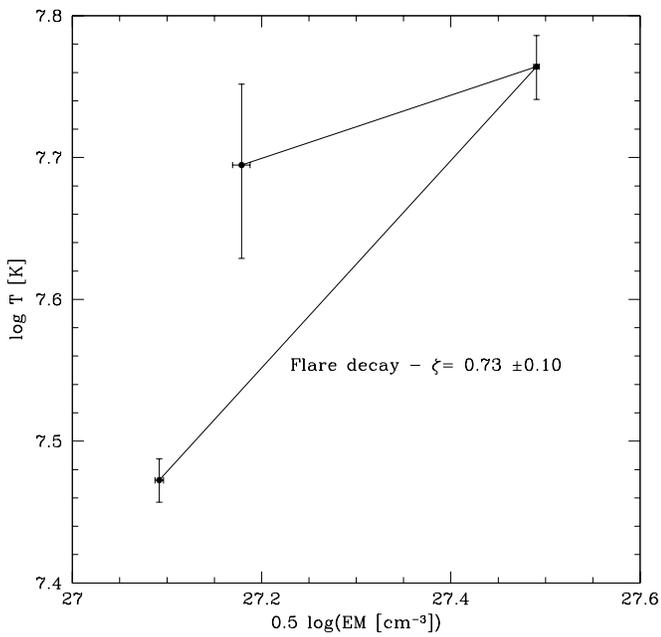, width=9.0cm}
        \caption{The evolution of the V827~Tau flare in the
            $\log T$ vs.\ $\log \sqrt{E\!M}$ plane.}
        \label{fig:zeta}
    \end{center}
\end{figure}

The $\log T$ vs.\ $\log \sqrt{E\!M}$ diagram for the V827~Tau flare is
  shown in Fig.~\ref{fig:zeta}. Applying the above formalism to the
  flare, we derive $\tau_{\rm LC} = 18.5$~ks and a slope $\zeta = 0.73
  \pm 0.10$.  Applying Eqs.~\ref{eq:fchi} and~\ref{eq:loop}, the
  resulting loop semi-length is $L = 2.5-3.5,R_\odot$. Given its
  bolometric luminosity and temperature ($L_{\rm bol} = 1.1 L_\odot$,
  $T_{\rm eff} = 4060$~K, as estimated by \citealp{blh+2002}), the
  radius of V827 Tau is $R \simeq 2\,R_\odot$, so the loop has a size
  comparable to the star itself.  Assuming for the loop a radius $r =
  0.1 L$, typical of solar events, from the emission measure at the
  flare maximum one derives an electron density $n_e \simeq 1.3 \times
  10^{11}$~cm$^{-3}$, which is typical of the values found in the
  intense flares of YSOs (\citealp{ffr+2005}). The corresponding
  equipartition magnetic field strength is $B \simeq 340$ G.

\begin{table*}[thbp]
  \begin{center}

  \caption{Best-fit spectral parameters of the flaring component in
    V827 Tau during flare episode (Obs. 1001 has been subdivided into
    2 segments). $Z = 0.3~Z_{\odot}$ (frozen); units are $E\!M_{53} =
    10^{53}$~cm$^{-3}$. The parameters of the first two
      components were frozen to values derived by fitting four spectra
      during the quiescent phase (to the values \nh\ $=0.09 \times
      10^{22}~{\rm cm^{-2}}$, $kT_1=0.36$~keV, $E\!M_1= 3.60 \times
      10^{53}$~cm$^{-3}$, $kT_2=1.16$, $E\!M_2=6.64 \times
      10^{53}$~cm$^{-3}$, and $Z=0.11~Z_{\odot}$) as described in the
      text.}
    \leavevmode
        \footnotesize
    \begin{tabular}{r|ccccc}
Obs. & $kT$ & $E\!M$ & $\chi^2$ & $P$ & Rate\\
\hline
~ &  keV & $E\!M_{53}$ & ~ & ~ & cts/s\\
\hline
1001A &  4.27 $\pm$ 0.60 &  22.78 $\pm$ 0.97 & 0.90 & 0.71 & 1.57 $\pm$ 0.10\\
1001B &  5.01 $\pm$ 0.26 &  95.75 $\pm$ 1.30 & 1.20 & 0.01 & 5.67 $\pm$ 0.09\\
1101 &   2.56 $\pm$ 0.09 &  15.26 $\pm$ 0.31 & 1.34 & 0.00 & 1.14 $\pm$ 0.02\\
    \end{tabular}
    \label{tab:v827_flare}
  \end{center}
\end{table*}

\subsection{V1075 Tau}

V1075 Tau is a K7 binary WTTS with a 2.43 d rotational period
(\citealp{bck+95}). In the 2000 observation the V1075~Tau count rate
decreased by a factor of $\simeq 2$ in about 30 ks. The FGM03 spectral
analysis showed that, while the temperature did not change
significantly, the absorption varied from $0.19\pm 0.03 \times
10^{22}~{\rm cm^{-2}}$, when the source was more intense, to $0.08\pm
0.03 \times 10^{22}~{\rm cm^{-2}}$ at the lower flux level.

\begin{figure}[!tbp]
  \begin{center} \leavevmode \epsfig{file=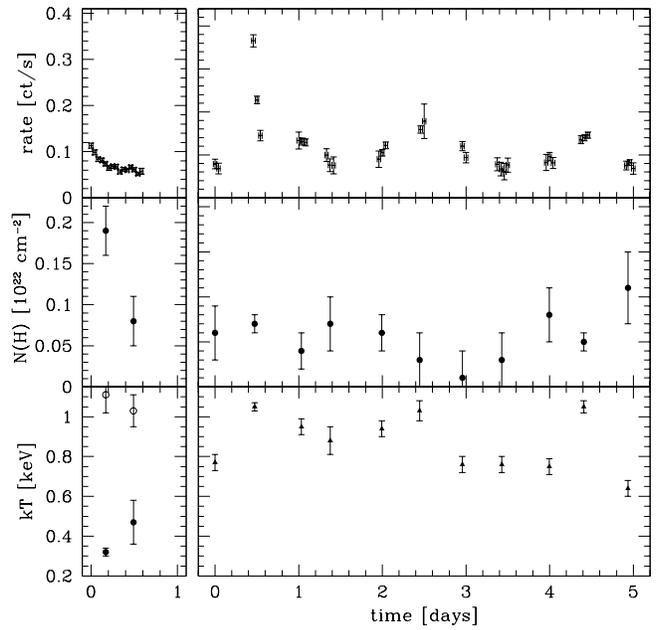, width=9.0cm}

    \caption{{\em Right panel -} From top to bottom: light curve of
      V1075 Tau over the 5 days of monitoring (bin-time 3600 s) and
      best-fit values of \nh\ and $kT$. {\em Left panel -} light curve
      and best-fit values of spectral parameters for V1075 from the
      \xmm\ observation of Sept.  2000 (All are PN data). An absorbed
      2T-plasma model was used for the spectral fits of the 2000
      data. All data are from PN.}
        
	\label{fig:v1075all}
    \end{center}
\end{figure}

The light curve of V1075 Tau is shown in Figure~\ref{fig:v1075all}.
Variability over a range of time scales is present, including a flare in
the second exposure and an apparent modulation on a time scale of a
couple of days.

\begin{figure}[!tbp]
        \begin{center} \leavevmode \epsfig{file=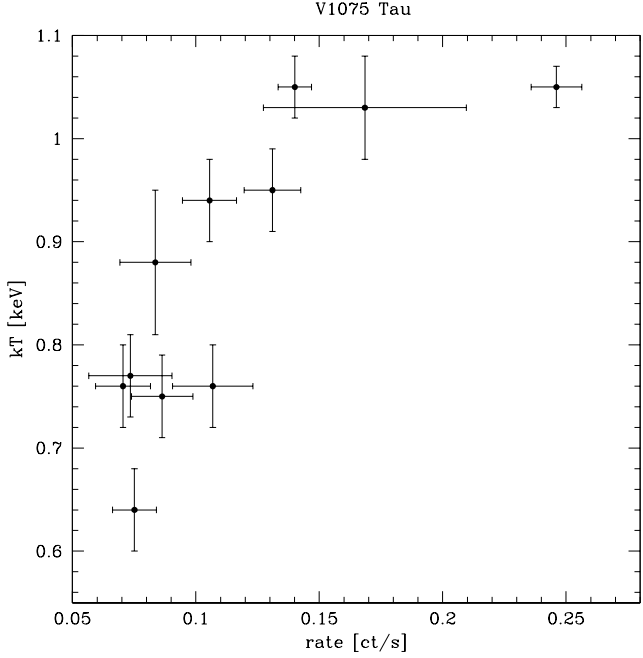, width=9.0cm}
        \caption{Count rate of V1075~Tau versus plasma temperature for
          the 11 exposures.} 
        \label{fig:v1075LcVkT}
    \end{center}
\end{figure}

The spectral parameters are reported in Table~\ref{tab:v1075_all11pn}.
Given the moderate statistics, the data could be satisfactorily fit
with a single temperature plasma. No changes in \nh\ are visible,
while $kT$ varies between 0.75 and 1.2 keV. Figure~\ref{fig:v1075LcVkT}
shows a scatter plot between the spectral temperature and the count
rate. The two are clearly correlated ($P=0.99993$ from a Wilcoxon
test, \citealp{wil45}), showing that the variability is more
likely intrinsic, rather than due to rotational modulation (which
would probably result in a 'gray' modulation, i.e.\ a
  spectral-independent change in the X-ray flux).

The spectral parameters for the merged spectrum (\nh$=0.10\pm 0.01
\times 10^{22}$~cm$^{-2}$, $kT_1 = 0.37\pm 0.02$, $kT_2=1.00\pm 0.02$
and $Z=0.19\pm 0.03$) are similar to the FGM03 values, as is the X-ray
luminosity ($L_{\rm X} = 2.0 \times 10^{30}$~\es).

\subsection{V710 Tau A \& B}

\begin{figure}[!tbp]
  \begin{center} \leavevmode \epsfig{file=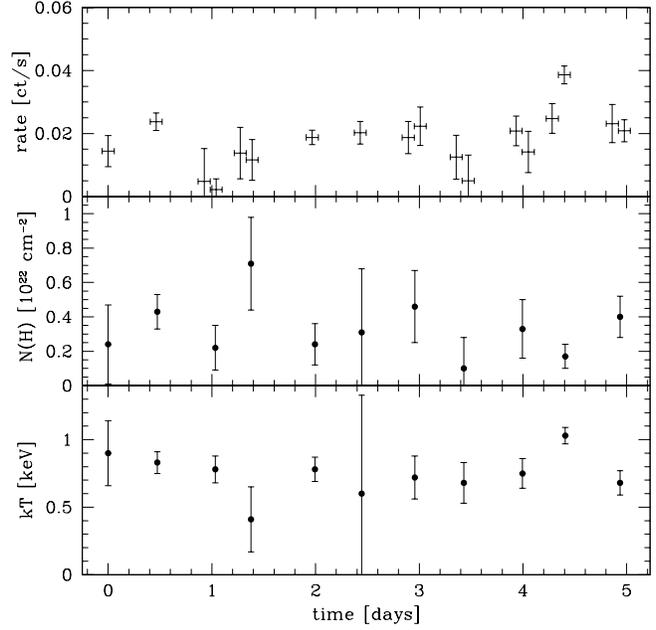, width=9.0cm}
    \caption{From top to bottom: light curve of V710 Tau over the
      5 days of monitoring (bin-time 3600 s, PN data) and best-fit
      values of \nh\ and $kT$ (PN, MOS1 and MOS2 data). PN data from
      \xmm\ observation of 2000 are not given for this source because
      they were found unreliable after reprocessing with more recent
      SAS pipeline (see text).}
    \label{fig:v710all}
  \end{center}
\end{figure}

V710 Tau is a CTTS+WTTS binary (\citealp{cfk+96}) with spectral types
M1 and M3 and masses 0.68~$M_{\sun}$ and 0.48~$M_{\sun}$
(\citealp{ja2003}). With a projected separation of 3.2 arcsec, the
system is not resolved by \xmm. Its light curve is shown in
Fig.~\ref{fig:v710all}. The system shows significant X-ray variability
with the source counts varying by more than a factor of two. FGM03
found no evidence of variability from V710 Tau during the 50 ks 2000
\xmm\ observations.

Due to the faintness of the source, PN and MOS data were fitted
simultaneously with a single temperature model;
Table~\ref{tab:v710_all11pnmos12} reports the spectral parameters.
The merged PN spectrum of V710 Tau has sufficient statistics for a 2T
fit. The spectral parameters (\nh$=0.19\pm 0.06 \times
10^{22}$~cm$^{-2}$, $kT_1 = 0.30\pm 0.053$ $kT_2=0.93\pm 0.06$,
$Z=0.2$ frozen) showed somewhat lower temperatures than in the 2000
observation ($kT_1=0.63\pm 0.05$, $kT_2 = 1.24\pm 0.10$, FGM03).
However, the SAS V6.0.0 reprocessing of the FGM03 data flags most of
the source photons as invalid (due to a row of hot pixels). The
reprocessed MOS2 data of V710 Tau are, on the other hand, `clean', and
the best-fit to the 2000 data gives \nh$=0.30\pm 0.08 \times
10^{22}$~cm$^{-2}$, $kT_1 = 0.38\pm 0.10$, $kT_2=1.12\pm 0.06$
($P=0.06$), similar to the values of the current observation.  The
source X-ray luminosity from the merged data is $0.6\times
10^{30}$~\es, a factor of 2 lower than the 2000 value ($1.3\times
10^{30}$~\es).

\subsection{XZ Tau}

XZ Tau is a binary CTTS with 0.3 arcsec separation (\citealp{hlz90}),
associated, together with HL Tau, with a complex set of bipolar jets
and Harbig Haro outflows (\citealp{mbs+90}).  The spectral types are
M2 and M3.5 for XZ Tau North and XZ Tau South, respectively
(\citealp{hk03}).  A photometric period of 2.6 days has been derived
by \cite{bck+95}. They interpret this period as due to rotational
modulation by a hot spot, 1500 K hotter than the photosphere and
covering 1.2\% of the stellar surface.
 
During the 2000 \xmm\ observation (FGM03), the X-ray count rate
increased by a factor of four in an approximately linear fashion over
50~ks. A time-resolved spectral analysis of the X-ray emission
resulted in significant spectral changes, in particular a decrease in
\nh\ from $1.06\times 10^{22}$~cm$^{-2}$ to $0.26\times
10^{22}$~cm$^{-2}$. The temperatures increased from $kT_1=0.14$~keV
and $kT_2=2.29$~keV to $kT_1=1.00$~keV and $kT_2=4.98$~keV (Table 5 in
FGM03). The average spectrum was described well by a very
low-metallicity plasma ($Z=0.007~Z_\odot$), with an X-ray luminosity
$L_{\rm X}=1.3 \times 10^{31}$~\es.

\begin{figure}[!tbp]
  \begin{center} \leavevmode \epsfig{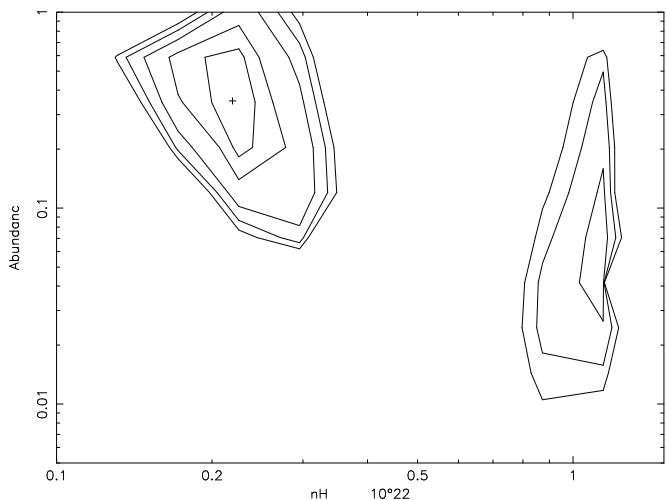}

\caption{Contours of $\chi^2$ as a function of the spectral parameters
\nh\ and $Z$ for the fit to the 2000 spectrum of XZ~Tau reprocessed
with the V6.0.0 software, during the first time interval (the 50 ks
observation was subdivided in three intervals). The $\chi^2$ 
space presents two well-separated minima that are both
acceptable solutions. The cross indicates the solution implying no
change in \nh.}

\label{fig:chi-landscape}
\end{center}
\end{figure}

We reprocessed the data with SAS V6.0.0 and re-analyzed them,
finding that the variation of \nh\ reported by FGM03 may be spurious.
The re-analysis shows that the spectral data for the first interval
can be equivalently fit with two different solutions, one with the
high \nh\ reported by FGM03 and another with a low \nh\ compatible
with the later intervals. The two solution spaces are separated
in the $\chi^2$ space well (as visible in
Fig.~\ref{fig:chi-landscape}) and both have an acceptable best-fit
probability. However, 'Occam razor' arguments lead us to prefer the
one in which \nh\ does not vary with respect to the rest. In fact,
well-separated minima in the $\chi^2$ space are also
present for the integrated 2000 spectrum.

The results of the spectral fits to the reprocessed 2000 data are
given in Table~\ref{tab:xztau_old} (replacing Table 5 of FGM03). 
  The only spectral parameter that varies significantly over the 50 ks
  is the emission measure of the hotter component, which increases by
  a factor of five, with no appreciable increase in temperature. A
  very-similar phenomenon -- a large "slow" increase in the X-ray flux
  without significant change in the plasma temperature -- was observed
  in the outbursting YSO V1674 Ori (\citealp{gko+2005}). The average
spectral parameters for the spectrum of the entire 50 ks observation
resulting from this re-analysis are $\nh=0.24 \pm 0.01$, $kT_1=0.83
\pm 0.02$~keV $kT_2=4.09\pm 0.34$~keV, and $Z=0.22\pm 0.08$ for a
corresponding intrinsic luminosity of $2.3 \times 10^{30}$~\es, a
factor of 6 lower than reported by FGM03, due to the change in \nh.

\begin{table*}[thbp]
  \begin{center} 
        \caption{Best-fit spectral parameters for XZ Tau during three
          consecutive time intervals of the 2000 \xmm\ observation.
          See the text for the details.}

    \leavevmode
    \begin{tabular}{ccccccccc}
Time interval & $N({\rm H})$ & $kT_1$ & $kT_2$ & $E\!M_1$ & $E\!M_2$ &
$Z$ & $\chi^2$ & $P$ \\
\hline
 ks & $10^{22}~{\rm cm^{-2}}$ & keV & keV & $10^{53}$ cm$^{-6}$ &
 $10^{53}$ cm$^{-6}$ & $Z_\odot$ & ~ & ~\\
\hline

0--20 & $0.22\pm 0.04$ & $0.84\pm 0.03$ & $4.31\pm 1.34 $  & $0.32 \pm 0.59$
& $0.49\pm 0.10$ & $0.40\pm 0.34$ & 1.33 & 0.06\\

20--40 & $0.22\pm 0.021$ & $0.82\pm 0.04$ & $4.81\pm 0.77 $  & $0.69\pm 0.83$
& $1.97\pm 0.57$ & $0.20\pm 0.12$ & 0.79 & 0.95\\

40--54 & $0.24\pm 0.02$ & $0.83\pm 0.06$ & $3.51\pm 0.38 $  & $0.62\pm 0.90$
& $2.46\pm 0.28$ & $0.24\pm 0.15$ & 0.94 & 0.63\\

\end{tabular}
    \label{tab:xztau_old}
  \end{center}
\end{table*}

\begin{figure}[!tbp]
  \begin{center} \leavevmode \epsfig{file=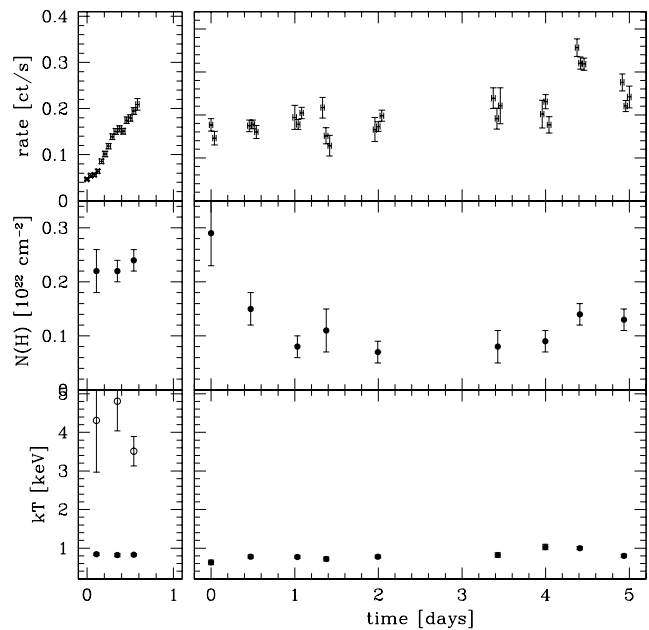, width=9.0cm}
    \caption{{\em Right panel -} From top to bottom: light curve
      of XZ Tau over the 5 days of monitoring (bin-time 3600 s) and
      best-fit values of \nh\ and $kT$ (Obs. 0701
      and Obs. 0801 have been). {\em Left panel -} light curve
      and best-fit values of spectral parameters for XZ Tau from the
      \xmm\ observation of Sept.  2000; for the spectral fits of the
      2000 data an absorbed 2T-plasma model was used. All data are
      from PN.}
    \label{fig:xzall}
  \end{center}
\end{figure}

The 2004 \xmm\ light curve of XZ~Tau is shown in Fig.~\ref{fig:xzall}.
The data points relative to Obs.\ 0701 and Obs.\ 0801 have been
omitted as they are strongly contaminated from the nearby HL~Tau, which
during this time is undergoing a large flare (see Sect.~\ref{sec:concl}).
 
Table~\ref{tab:xz_all11pn} summarizes the results of the spectral
analysis of the PN data for the individual exposures; these were
satisfactorily fit by an absorbed 1T plasma model. The spectral fits
to the data for observations 0401 and 1101 can be improved by using a
two-temperature plasma model (with larger error bars on the best-fit
parameters); to allow a uniform comparison of the time variability we
use the results of the 1T spectral fits.

Figure~\ref{fig:xzall} shows the best-fit values of \nh\ and $kT$.
While variability of a factor of $\simeq 2$ in the count rate is
present, no significant spectral variations are detected.  A spectral
fit to the merged data for XZ Tau (with the exclusion of Obs.  0701,
0801, and 0901, to prevent contamination from the HL Tau flare) yields
\nh$=0.19 \pm 0.03 \times 10^{22}$~cm$^{-2}$, $kT_1=0.34\pm 0.03$,
$kT_2=1.22\pm 0.07$, and $Z=0.14\pm 0.04$ ($P=0.02$, $\Delta E =
0.3-3.0$~keV) for a corresponding luminosity of $1.3\times 10^{30}$\es
(a factor of 2 lower than in 2000).

\begin{figure}[!tbp]
  \begin{center} \leavevmode \epsfig{file=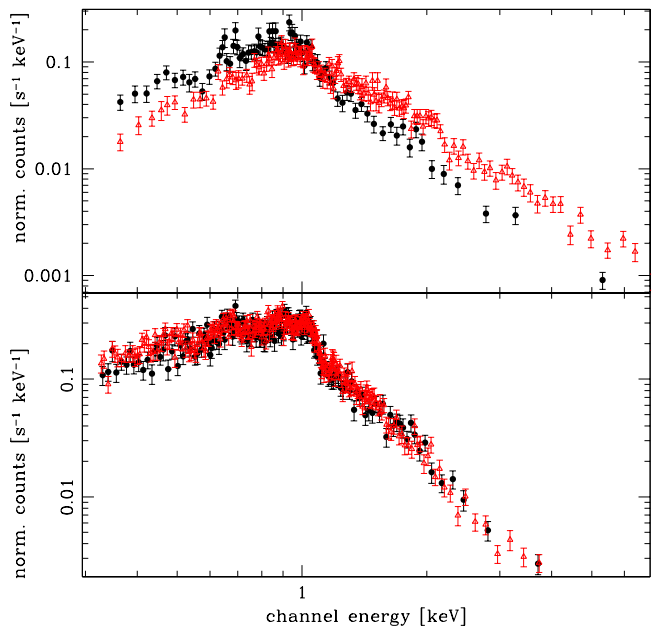, width=9.0cm}
    \caption{Spectra of XZ~Tau (top panel) and V826~Tau (bottom
      panel) from the merged PN data from the 2004 monitoring campaign
      (filled circles) and from the 2000 observations (open
      triangles). While the V826~Tau spectra are almost identical, the
      hard component present in 2000 in XZ~Tau is no longer visible.}
    \label{fig:xz_comp}
  \end{center}
\end{figure}

The plasma temperature during the 2004 observation is significantly
lower than in 2000, with no evidence of the hot ($T\simeq 4$ keV)
component clearly present in the 2000 data. Visual inspection of the
two spectra (top panel of Fig.~\ref{fig:xz_comp}, which plots both
spectra reprocessed with the standard SAS V6.0.0 pipeline) shows that
indeed the high-energy tail clearly visible in the 2000 spectrum has
weakened significantly. For comparison, the bottom panel of
Fig.~\ref{fig:xz_comp} shows the 2000 and 2004 spectra of V826~Tau,
for which no differences are visible, allowing us to exclude e.g.\ 
problems in the background filtering.

XZ~Tau is the only star in the sample showing substantial spectral
changes from 2000 to 2004. In 2000 XZ~Tau had a hotter X-ray spectrum
than in either 2001 or in 2004: the 2000 spectrum indicates $kT_1=0.83
\pm 0.02$~keV and $kT_2=4.09\pm 0.34$~keV, while the present (2004)
data indicate $kT_1=0.3-0.7$~keV and $kT_2=1.0-1.5$~keV, which are
very similar to the values derived from the July 2001 \chandra\
observation (FGM03), $kT_1 = 0.65 \pm 0.03$ and $kT_2 = 1.56 \pm
0.12$. The light curve of the 2000 observation shows no obvious
evidence of flaring (which would justify the higher temperature); slow
variability is present, starting from a lower value than any of the
present observations and rising to a value similar to the 2004 one.

Ground-based $UVBR$ photometry of XZ~Tau from 1962 to 1993
detected variations of $\ge 2$ mag (\citealp{hhg+94}), integrated on
both components of the binary system. \cite{cdr04} monitored XZ Tau
and its outflows from 1995 till 2001 using HST, which resolves the
binary. XZ Tau South ($R \simeq 13.5$) displays moderate variability
($\Delta R \le 0.3$ mag), while XZ Tau North (the suspected source of
the outflow) displays dramatic variations: in Jan. 1995 its
magnitude was $R = 14.93$, fading by about 1 mag until 1998 and
thereafter brightening by 3 mag, so that by Feb.\ 2001 XZ Tau North
was actually the brighter star (also visible in Fig.~1 of
\citealp{cdr04}). This behavior suggests that XZ Tau North is an EXor.
EXors, named after their prototype EX Lupi, are a loosely defined
  class of eruptive CTTS that periodically undergo outbursts from the
  UV to the optical.  Although increases by several magnitudes with
  rise times of up to a few years have been recorded
  (\citealp{her89b}), the changes in these YSOs are not as extreme as
in FU Ori stars (EXor spectra during outburst, for example, continue
to resemble the ones of T Tauri stars). This phenomenon is thought to
be due to major increases in the underlying disk accretion rate; 
  however, the number of known EXors is relatively small and the class
  remains poorly defined.

\begin{figure}[!tbp]
 \begin{center} \leavevmode \epsfig{file=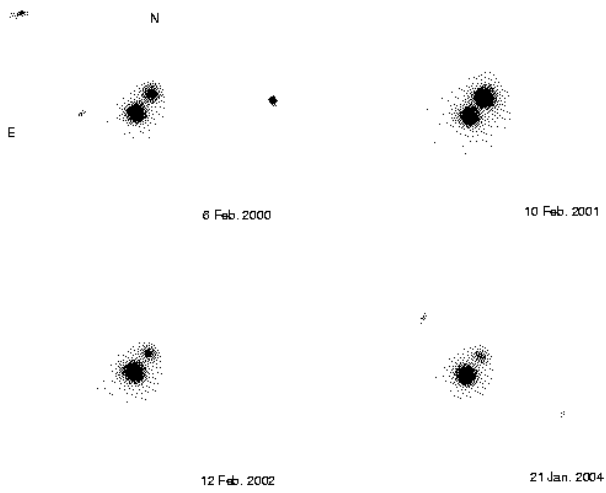, width=9.0cm}
        \caption{A mosaic of HST images of the XZ Tau binary taken
          between 2000 and 2004 (field size $4'' \times 3''$
            centered on XZ Tau, at 04:31:40.07 $+$18:13:57.1). While
          during the outburst episode of 2001 XZ Tau N was the
          brightest component, 1 year later it had faded back to being
          the fainter component, and by early 2004 (when the \xmm\ 
          observations discussed here have been taken) the S component
          is much stronger. The 2000, 2001, and 2002 images are from
          the WFPC2 camera, in the F675W filter, while the 2004 is
          from the ACS camera in the FR656N filter.} \label{fig:xzhst}
      \end{center}
\end{figure}

While the HST data presented by \cite{cdr04} extend only to Feb. 2001,
a number of more recent HST observations are present in the public
archive, which can help to determine whether the outburst that was
ongoing at the time of the 2000 observation has lasted into the 2001
\chandra\ and 2004 \xmm\ observations. We have located a number of
unpublished observations in the HST archive, which we extracted
and inspected. While a full photometric analysis of these observations
is beyond the scope of the present work, even simple visual inspection
of the images allows assessment of the luminosity of XZ Tau N (the
component found to strongly vary by \citealp{cdr04}) compared to the
more stable XZ Tau S. We have chosen four HST observations (all taken
in red filters) to be as close as possible in time to the X-ray
observations. A zoom centered on the resolved XZ Tau binary is shown
in Fig.~\ref{fig:xzhst}. The first three observations are all from the
WFPC2 camera, in the F675W filter, while the last (2004) observation
is taken with the ACS camera, with the FR656N filter. The \xmm\
observation of FGM03 took place in Sept. 2000 between the HST
observation of Feb. 2000, when the brightening of XZ Tau N was already
going on (\citealp{cdr04}), and the HST observation of Feb. 2001,
when the outburst was clearly visible.  The outburst, however, did not
appear to be long lasting: one year later (Feb.  2002) the outburst
was already finished, with XZ Tau N back to being fainter than the S
component. In early 2004 it was, if anything, even fainter than
the S component.  The \chandra\ observation of July 2001 fell
between the early 2001 and early 2002 HST observations, and the present
\xmm\ campaign (4--9 Mar.\ 2004) is very close in time to the 2004 HST
observation.

While the X-ray and optical observations are not simultaneous and
  the binary system is not resolved by \xmm, the temporal proximity
of the optical outburst of XZ Tau N and of the X-ray spectral
hardening (as well as of the peculiar variability observed in 2000) is
very suggestive of a connection between the two phenomena.  Such
  a connection has been well-documented for V1647 Ori by
  \cite{krg+2004}, who reports a significant increase in spectral
  hardness near the peak of the optical outburst, as well as a surge
  of X-ray flux.  Moreover, as already mentioned, the outbursting
  source V1647 Ori also displays the type of `slow'  variability
  observed for XZ Tau in 2000 ({\citealp{gko+2005}).

\subsection{HL Tau}
\label{sec:hltau}

HL Tau, at about 24 arcsec from XZ Tau, is a K7 embedded young stellar
object, often considered a prototype very young solar-mass star
(0.5$-$0.7~$M_{\sun}$), with a circumstellar disk that resembles the
solar nebula at the early stages of planet formation
(\citealp{mhf99}). Together with XZ~Tau, it is associated with bipolar
jets and Herbig-Haro outflows (\citealp{mbs+90}).

Although initially classified as a CTTS, it is now thought to be in a
transition phase from protostar (Class I object) to CTTS. An infalling
envelope around the source was found by \cite{hom93}, who reported
evidence of non-steady accretion. \cite{crn+97} found that to
reproduce the observed SED, the central source in HL Tau is required
to be a very young ($\sim 10^5$~yr) PMS surrounded by an active
accretion disk and accreting at the rate of $5\times
10^{-6}~M_{\sun}$~yr$^{-1}$.

Ground-based photometry of HL Tau from 1973 to 1993 has shown
variations of 0.5 to 2 mag in $UVBR$ (\citealp{hhg+94}).  During the
2000 \xmm\ 50ks observations HL~Tau did not display significant
variability, while in the 2001 80 ks \chandra\ observation
(\citealp{bfr2003}), it underwent a small short duration flare (Fig.~6
in FGM03). As shown in Figure~\ref{fig:hlall} in the present observation,
HL Tau underwent a large ($\times~ 25$ in count rate) flare
which decays over about two days.

The best-fit spectral parameters are reported in
Table~\ref{tab:hl_all11pn}. Given the low statistics of individual
spectra, the metal abundance was frozen to $Z=0.6~Z_{\odot}$, the
value determined in 2000 by FGM03. Outside of the flare the spectral
parameters do not vary significantly and their average values (see
below) are similar to the 2000 values (FGM03), as is the X-ray
luminosity ($1.5\times 10^{30}$~\es).

\begin{figure}[!tbp]
  \begin{center} \leavevmode \epsfig{file=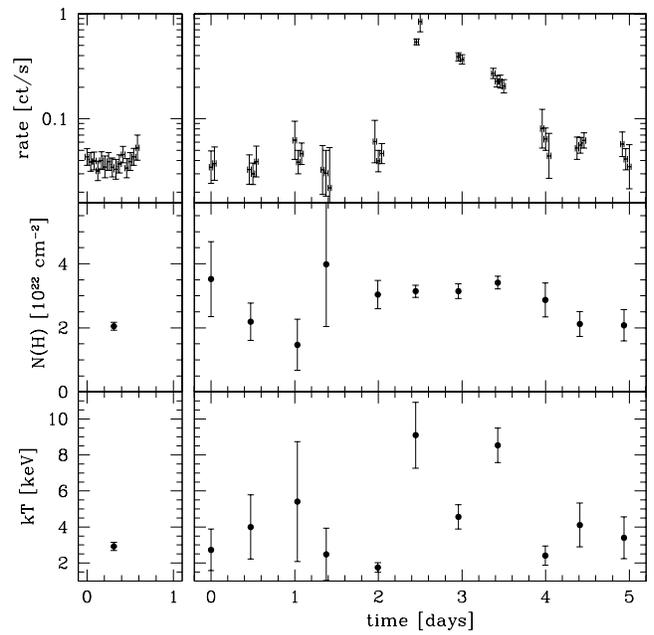, width=9.0cm}
    \caption{{\em Right panel -} From top to bottom: light curve
      of HL Tau over the 5 days of monitoring (bin-time 3600 s) and
      best-fit values of \nh\ and $kT$. {\em Left panel -} light curve
      and best-fit values of spectral parameters for HL Tau from the
      \xmm\ observation of Sept.  2000.}
    \label{fig:hlall}
  \end{center}
\end{figure}

\begin{table*}[thbp]
  \begin{center}
  \caption{Best-fit spectral parameters of the flaring component in HL
    Tau. Units are $N_{22} = 10^{22}~{\rm cm^{-2}}$, $E\!M_{53} =
    10^{53}$~cm$^{-3}$ and $F_{-13}= 10^{-13}$~\ecms. }
 
  \leavevmode
  \footnotesize
  \begin{tabular}{r|ccccccc}
    Obs. &  $N({\rm H})$ & $kT$ & $E\!M$ & $Z$ & $\chi^2$ & $P$ & Rate\\
    \hline
    ~ &   $N_{22}$ & keV & $E\!M_{53}$ & $Z_{\odot}$ & ~ & ~ & cts/s\\
    \hline
    0701 &  3.40 $\pm$ 0.27 & 8.79 $\pm$ 2.50 & 14.01 $\pm$ 1.87 & 0.48 $\pm$ 0.19 & 0.75 & 0.90 & 0.48 $\pm$ 0.01\\
    0801 &  3.73 $\pm$ 0.33 & 3.86 $\pm$ 0.62 & 15.88 $\pm$ 3.08 & 0.26$\pm$ 0.11 & 0.87 & 0.73 & 0.35 $\pm$ 0.01\\
    0901 &  4.29 $\pm$ 0.35 & 7.02 $\pm$ 1.51 & 9.55 $\pm$ 1.57 & 0.27$\pm$ 0.11 & 0.88 & 0.71 & 0.26 $\pm$ 0.01\\
  \end{tabular}
  \label{tab:hl_flare}
\end{center}
\end{table*}

To derive the flaring emission spectral parameters, we applied the
same procedure as for V827~Tau. The quiescent spectral parameters are
\nh $=2.43 \times 10^{22}~{\rm cm^{-2}}$, $kT=3.10$~keV, $E\!M= 1.40
\times 10^{53}$~cm$^{-3}$, and $Z=0.6~Z_{\odot}$, and the resulting
spectral parameters for the flaring component are listed in
Table~\ref{tab:hl_flare}.  The event shows a peculiar evolution, with
a monotonically decaying light curve associated with a highly
irregular temperature evolution.  The temperature has two well-defined
peaks above 7 keV separated by a deep minimum at about 3 keV.  The
temperature evolution suggests that this is the combination of two
flares, probably physically related to each other but occurring in
independent coronal structures. This kind of evolution has been
predicted by modeling two independent flares by \cite*{rgp+2004}.

In order to support this hypothesis, we  modeled the event by
combining two flares computed with detailed hydrodynamic
modeling of plasma confined in a coronal loop. The light curve decay
time suggests long flaring structures (e.g.\ \citealp{srj+91}), so that
we considered each model flare to be identical to the one used to
describe one of the flares observed during the COUP campaign 
(\citealp{ffr+2005}) in detail. We assumed that each flare occurs in a coronal
loop with a constant cross-section and half-length $L = 10^{12}$ cm,
symmetric around the loop apex. Both flares were triggered by injecting
a heat pulse in the loop, which was initially at a temperature of
$\simeq 20$ MK. This heat pulse is symmetrically deposited at the loop
footpoints with a Gaussian spatial distribution of intensity 10 erg
cm$^{-3}$ s$^{-1}$ and width $10^{10}$ cm (1/100 of the loop
half-length). After 20 ks the heat pulse was switched off completely.
From the evolution of the plasma density and temperature along the
loop computed with the Palermo-Harvard hydrodynamic loop model
(\citealp{psv+82}, \citealp{bpr+97}), we synthesized the corresponding EPIC
spectra of the loop throughout the flare, deriving a light curve and
the evolution of temperature.

To model the HL~Tau event we duplicated, the resulting light curve and
temperature evolution with a time shift. The two flares are identical
flares, except for a normalization factor, which represents the loop
cross-section and does not enter explicitly in the hydrodynamic
modeling. The second flare has a normalization factor of 0.3 (i.e.\ a
correspondingly smaller cross-section) and it starts 60 ks after the
first. We summed the resulting two asynchronous sequences of flare
spectra and obtained a single sequence of spectra, which we integrated to
derive a single light curve and fit with single temperature EPIC model
spectra. Figure~\ref{fig:hlflare} shows the resulting light curve and
temperature evolution as compared to those obtained from the data.
The model temperatures in the first flare are somewhat higher
  than the observed one, but the main flare characteristics, i.e.\ the
  monotonic light curve and the temperature dip, are reproduced
  by the double-flare model well -- although the model is not unique.

Since no constraint can be derived from the data, each of the two
flares was modeled with no significant residual heating present
during the flare decay (e.g.\ \citealp{rbp+97}). Such modeling implies
very large flaring structures, similar to the ones found in ONC YSOs
by \cite{ffr+2005} -- where the data allowed investigation of the
presence of sustained heating. Such large structures, with $L \simeq
5\,R_*$, have only been found in YSOs, and were interpreted by
\cite{ffr+2005} as linking the star to the accretion disk, i.e.\ as
being the magnetic structures supporting the magnetospheric accretion.
In addition to the evidence from the ONC YSOs, HL~Tau is the first
Taurus YSO in which such large flaring structures have been detected.
We cannot a priori exclude that shorter loops with sustained heating
in the decay may also reproduce the features of this flare. However,
the long delay of the model flares required to reproduce the
distant temperature peaks suggests that very large structures must be
involved in the flare (\citealp{rgp+2004}).

\begin{figure}[!tbp]
  \begin{center} \leavevmode \epsfig{file=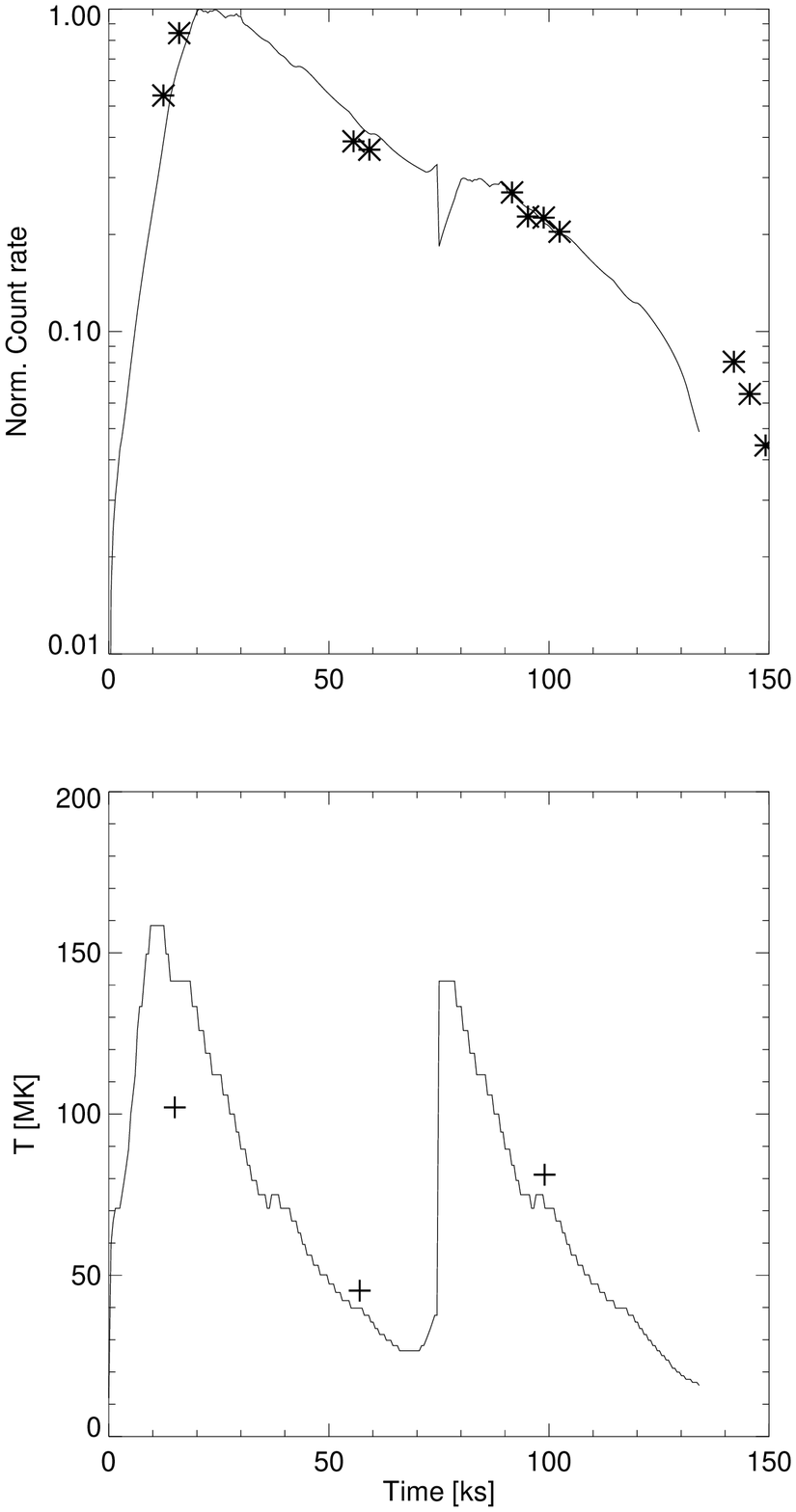, width=9.0cm}
    \caption{{\em Top panel -} The observed light curve for HL~Tau
    (asterisks), together with the light curve predicted by the flare
    modeling described in the text. {\em Bottom panel -} The observed
    temperature evolution of the HL~Tau flare, together with the
    double-peaked temperature evolution predicted by the same flare
    modeling. The model is obtained by combining two models of $2
    \times 10^{12}$ cm-long flaring loops ({\it dotted and dashed
    lines}). The two models are identical, except for the
    cross-section of the flaring loops and for the start time (60 ks
    delay).}  \label{fig:hlflare} \end{center}
\end{figure}

\subsection{HD\,285845}

Unlike the other stars in the present study, HD\,285845 (an active
binary system) is not a member of the star-forming region, on the
basis of its radial velocity and proper motion
(\citealp{wbm+88}). The primary spectral type is G8, and \cite{shw98}
report a separation of 73 mas and a magnitude difference of 1.19 mag.
 
In the 2000 observation, HD\,285845 showed significant variability,
and similar behavior is present in the present data set
(Fig.~\ref{fig:hdall}).  Table~\ref{tab:hd_all11pn} summarizes the
best-fit spectral parameters. Notwithstanding the significant
count-rate variability, the spectral parameters show little variation
in time.

\begin{figure}[!tbp]
  \begin{center} \leavevmode \epsfig{file=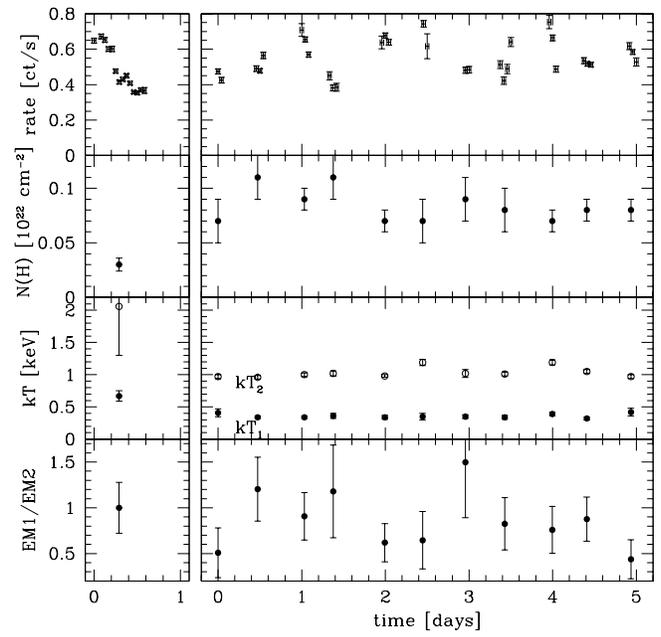, width=9.0cm}
    \caption{{\em Right panel -} From top to bottom: light curve
      of HD285845 over the 5 days of monitoring (bin-time 3600 s),
      best-fit values of \nh\, $kT_1$, $kT_2$ and $E\!M_1/E\!M_2$
      ratio.  {\em Left panel -} light curve and best-fit values of
      spectral parameters from the \xmm\ observation of Sept.  2000
      (All are PN data).}
    \label{fig:hdall}
  \end{center}
\end{figure}

\section{Temporal variability}
\label{sec:vari}

All the stars in the sample show significant variability, although
with different characteristics. To quantify this variability we 
computed the normalized cumulative distributions of the amplitude
variability. These represent the fraction of time that a source spends
in a state with the flux larger than a given value, expressed in terms of
a given a normalization value, which can be the minimum count rate,
the median count rate, etc. For the present sample we took as
normalization value the count rate above which a source spends
  90\% of the time; this is less sensitive to noise fluctuations than
  the real minimum.

Figure~\ref{fig:intAll} shows the distribution for the stars in our
sample (except V710~Tau for which the count rate uncertainties are too
large). Some sources (V826~Tau, HD285845 and XZ~Tau) show mostly
low-amplitude variability, in which less than 30\% of the time is
spent in a state 1.3$-$1.5 above the minimum. On the other hand,
V827~Tau and HL~Tau spend more than 60\% of the time in such a state,
while V1075~Tau shows intermediate behavior.  The differences between
V827~Tau and HL~Tau are not due to the large flares present in their
light curves, as the flare points affect only the 10 bins relative to
the highest count rate, while the low-variability tails of the
distributions are also significantly different.

Kolmogorov-Smirnov tests for the cumulative distributions of various
pairs of sources also indicate the presence of the different
behaviors. While this would suggest differences in the X-ray emission
processes, the two groups are heterogeneous, and both contain CTTS and
WTTS, so that a simple interpretation in terms of e.g. accreting vs.
non-accreting sources is not possible.

\begin{figure}[!tbp]
  \begin{center} \leavevmode \epsfig{file=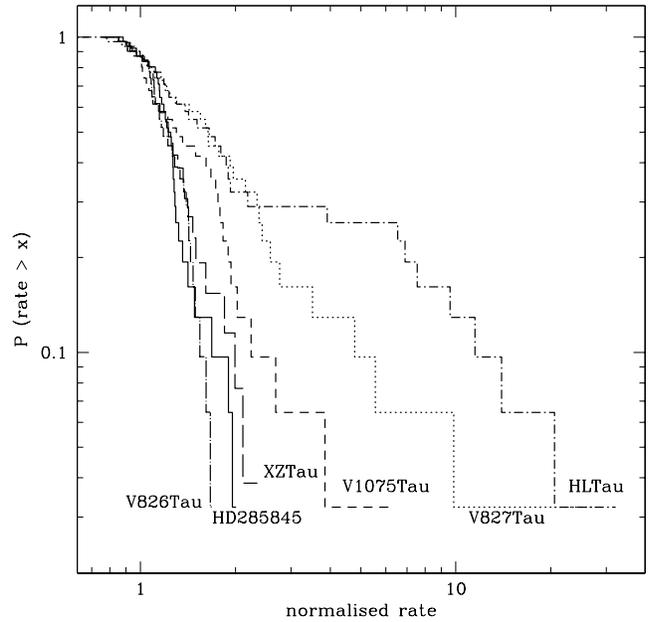, width=9.0cm}
    \caption{Normalized cumulative count-rate distribution for all
      the star considered apart from V710~Tau.}
    \label{fig:intAll}
  \end{center}
\end{figure}

\section{Discussion}
\label{sec:disc}

The X-ray monitoring observations of the L1551 star-forming complex
discussed here, in conjunction with the 2000 \xmm\ observations
(FGM03) and the 2001 \chandra\ observations (\citealp{bfr2003}), has
allowed us to probe the X-ray intensity and spectral variability from 
six YSOs (both CTTS and WTTS), on a variety of time scales, from
hours to days to years.

Most stars in the sample show X-ray variability at an amplitude of a
factor of $\simeq 2$ on time scales of a few days. In the WTTS V1075
Tau, the source count rate varies on time scales similar to its
optically derived rotational period, which would be compatible with
rotational modulation (as observed in YSO in the ONC,
\citealp{fms+2005b}), but the significant correlation with spectral
variability would appear to point to intrinsic variations. Plasma
temperature variations of the order of 20\% on similar time scales are
also present, and in some cases they appear to be correlated with the
intensity (e.g.\ for V1075 Tau).

Again XZ~Tau emerges as a peculiar source. While the short-term (day
scale) spectral variability of XZ~Tau is not linked to variations in
the absorbing column density, as originally speculated by FGM03 --
reprocessing and reanalysing the original 2000 \xmm\ data showing
that the variation reported by FGM03 was probably spurious -- this is the
only source in the sample to show significant long-term changes in its
spectrum. 

The spectrum of XZ Tau observed with \xmm\ in 2000 had a significant
hard component ($T \simeq 4$ keV), which was not visible in the 2001
\chandra\ data and is not present in the 2004 \xmm\ data and both are
compatible with a hard component of at most 1.5 keV. The difference
between the two spectral states is very well visible in
Fig.~\ref{fig:xz_comp}. The X-ray spectral hardening takes place close
in time to the strong optical outburst of the N component of the
binary. While the data by \cite{cdr04}, who reported the outburst,
only cover its beginning, our own inspection of new HST data indicates
that the outburst was not long-lived, with the source returning
  to its state fainter than XZ Tau S by early 2002, one year after
the early 2001 peak.

Finally, while (Fig.~\ref{fig:xzall}) the count rate varied smoothly
between 0.04 and 0.22 cts/sec during the 2000 observation, during the
2004 campaign it never decreased below 0.12 cts/sec (with a peak value
of $\simeq 0.35$ cts/sec), again showing the source to be in a
different state.

The binary system XZ Tau is not resolved in the X-ray, therefore
  one cannot exclude that XZ Tau S may have been the dominant X-ray
  source, at one or all the epochs at which it was observed.
  Nevertheless the observed phenomena are very reminiscent of the
  spectral hardening and variability observed in the outbursting
  source, V1647 Ori (\citealp{krg+2004}; \citealp{gko+2005}), so that it
  seems reasonable to interpret the X-ray spectral changes and
  variability of XZ Tau as connected with the outburst of XZ Tau N.
The potential physical mechanism causing the significantly
  harder X-ray spectrum (with a somewhat lower emission level) during
the optical outburst is difficult to assess. While the optical
outburst is naturally interpreted as an accretion outburst (with the
optical luminosity due to increased luminosity of the accretion
shock), why this should cause a hot plasma component to appear in the
star is not clear. Indeed, accreting YSOs in the COUP sample are
statistically less X-ray luminous than non-accreting ones
(\citealp{pkf+2005}) and less prone to show flares from large
magnetic structures (\citealp{ffr+2005}).

The temperature of the hot component observed in XZ Tau in 2000 is
such that it cannot be due to simple accretion-driven emission, as the
shock temperature in a low-mass T Tau star is too low . Its early M
spectral type corresponds, at this age, to a typical mass $M \simeq
0.5\,M_\odot$ and to a radius $R \simeq R_\odot$, resulting
(\citealp{cg98}) in a peak shock temperature $T\simeq 0.2$ keV.

Nevertheless, a heating of the plasma during an accretion outburst in
XZ Tau N would be consistent with the observed high X-ray plasma
temperatures in accreting objects, like Class I sources. In a study of
the $\rho$ Oph region, \cite{int+2003} find that Class I sources have
a higher $ \langle kT\rangle$ (sometimes exceeding 5~keV) on
average than Class II-III sources. This is confirmed by the study of
$\rho$ Oph by \cite{ogm2004} that finds an evolutionary trend from
Class I sources showing higher temperature and larger absorption to
Class II and III sources showing lower temperatures and smaller
extinction. In the present sample the only star with a consistently
high plasma temperature ($kT =2.5-4$~keV) is HL~Tau, a PMS in the
transition phase between Class I and Class II, which is still
undergoing substantial accretion.

Apart from XZ~Tau, no significant variations in the activity level of
our target YSOs on time scales of four years is observed. While
evidence of solar-like cyclical variation is accumulating
(\citealp{fmb+2004}; \citealp{rsf2005}) for low- and
intermediate-activity stars, long-term variations (whether cyclical or
not) in the activity levels of high-activity stars (including YSOs) do
not seem to be present.

Although the data do not allow a fully detailed modeling, the
double-temperature peaked flaring event observed in HL~Tau most likely
originates in very large magnetic structures, similar to the ones
observed in ONC YSOs. Such structures, postulated by the
magnetospheric accretion model, probably link the stellar photosphere to
the inner rim of the accretion disk.  In the ONC sample only objects
that are currently not accreting were shown to have very large
flaring magnetic structures (\citealp{ffr+2005}). The only active
accretor present in the sample showed flares confined to small
structures, similar to normal coronal flares. HL~Tau (with the caveat
linked to the limited spectral information available) would provide a
counter-example to the scheme observed in the ONC, being the first
strongly accreting YSO with large flaring magnetic structures.

\section{Conclusions}
\label{sec:concl}

The present monitoring campaign has allowed us to study the luminosity
and spectral variability of the X-ray emission from a number of YSO
over a range of time intervals, sampling in particular the variability
over a 5-day span, as well as over 4 years.

While most YSOs show a remarkably constant level of activity over the
4-year span sampled, XZ~Tau has shown significant spectral
variability apparently in conjunction with an optical outburst of
  the N component of the binary system, which took place in 2000.
The role of accretion in the X-ray of YSOs is a matter of current
debate, with a few spectral observations indicating differences in the
density and UV environment of the plasma in a few accreting YSOs, and
with a number of authors speculating on the possible relationship between
accretion and coronal heating. \cite{krg+2004} and
  \cite{gko+2005} report on spectral hardening and enhanced
  variability in V1647 Ori in conjunction with a well-documented
  optical/IR outburst, which provides evidence that strongly enhanced
  high-energy emission can occur as a consequence of high accretion
  rates. The significant spectral changes detected here supply similar
  evidence.

\begin{acknowledgements}
  
  B. Silva acknowledges support by grant SFRH/BEST/9534/2002 from FCT
  (Funda{\c c}ao para a Ciencia e a Tecnologia). We thank the
  referee Joel Kastner for his very useful comments that have helped
  improve  the paper.

\end{acknowledgements}

\begin{table*}[!pthb]
  \begin{center}
    \caption{Best-fit spectral parameters of the PN data for the
      eleven exposures of V826~Tau. $E\!M$ is the emission
        measure, $P$ the null-hypothesis probability of the fit,and
        $F_{\rm X}$ Intr. the "unabsorbed" X-ray flux. Units are
      $N_{22} = 10^{22}~{\rm cm^{-2}}$, $E\!M_{53} =
      10^{53}$~cm$^{-3}$, and $F_{-13}= 10^{-13}$~\ecms. The
      metallicity was frozen at $Z=0.17~Z_{\sun}$. The spectral
        fits were carried out in the energy range 0.3--7.5~keV.} 

    \leavevmode
    \footnotesize
    \begin{tabular}{r|ccccccccc}
      Obs. & $N({\rm H})$ & $kT_1$ & $E\!M_1$ & $kT_2$ & $E\!M_2$ & $\chi^2$ & $P$ & $F_{\rm X}$ & $F_{\rm X}$ Intr.\\
      \hline
      ~ &  $N_{22}$ & keV & $E\!M_{53}$ & keV &  $E\!M_{53}$ & ~ & ~ & $F_{-13}$  & $F_{-13}$ \\
      \hline

0201 & 0.17 $\pm$ 0.04 & 0.28 $\pm$ 0.02 & 6.51 $\pm$ 3.74 & 1.07 $\pm$ 0.06 & 3.74 $\pm$ 0.68 & 1.01 & 0.45 & 11.52 & 24.51\\
0301 & 0.06 $\pm$ 0.03 & 0.39 $\pm$ 0.06 & 1.19 $\pm$ 0.83 & 1.04 $\pm$ 0.05 & 1.65 $\pm$ 0.32 & 1.16 & 0.19 & 6.33 & 8.30\\
0401 & 0.10 $\pm$ 0.03 & 0.35 $\pm$ 0.04 & 2.06 $\pm$ 1.23 & 1.19 $\pm$ 0.06 & 2.46 $\pm$ 0.36 & 1 & 0.47 & 8.34 & 12.72\\
0501 & 0.15 $\pm$ 0.05 & 0.26 $\pm$ 0.02 & 6.04 $\pm$ 4.84 & 0.89 $\pm$ 0.07 & 3.49 $\pm$ 0.81 & 1.01 & 0.42 & 10.60 & 22.20\\
0601 & 0.10 $\pm$ 0.02 & 0.31 $\pm$ 0.02 & 3.51 $\pm$ 1.49 & 1.2 $\pm$ 0.04 & 4.88 $\pm$ 0.41 & 1.16 & 0.12 & 15.72 & 23.66\\
0701 & 0.15 $\pm$ 0.06 & 0.29 $\pm$ 0.04 & 3.70 $\pm$ 3.86 & 0.97 $\pm$ 0.08 & 2.62 $\pm$ 0.78 & 0.49 & 0.96 & 7.90 & 15.69\\
0801 & 0.01 $\pm$ 0.03 & 0.44 $\pm$ 0.29 & 0.55 $\pm$ 0.74 & 1 $\pm$ 0.07 & 2.25 $\pm$ 0.70 & 1.02 & 0.44 & 8.81 & 9.06\\
0901 & 0.13 $\pm$ 0.04 & 0.26 $\pm$ 0.03 & 4.23 $\pm$ 2.80 & 0.83 $\pm$ 0.05 & 3.18 $\pm$ 0.85 & 1 & 0.51 & 9.65 & 17.98\\
1001 & 0.11 $\pm$ 0.04 & 0.31 $\pm$ 0.03 & 3.06 $\pm$ 2.07 & 0.98 $\pm$ 0.06 & 2.73 $\pm$ 0.56 & 0.95 & 0.62 & 9.02 & 15.15\\
1101 & 0.15 $\pm$ 0.03 & 0.28 $\pm$ 0.02 & 4.50 $\pm$ 2.18 & 0.96 $\pm$ 0.04 & 3.18 $\pm$ 0.42 & 1.06 & 0.34 & 9.44 & 18.80\\
1201 & 0.11 $\pm$ 0.03 & 0.27 $\pm$ 0.03 & 2.83 $\pm$ 1.80 & 0.94 $\pm$ 0.05 & 2.93 $\pm$ 0.45 & 0.87 & 0.79 & 8.92 & 14.92\\

    \end{tabular}
    \label{tab:v826_all11pn}
  \end{center}
\end{table*}

\begin{table*}[thbp]
  \begin{center}

  \caption{Best-fit spectral parameters of the PN data for the eleven
  exposures of V827~Tau. $E\!M$ is the emission measure, $P$ the
  null-hypothesis probability of the fit, and $F_{\rm X}$ Intr. the
  "unabsorbed" X-ray flux.} Units are $N_{22} = 10^{22}~{\rm
  cm^{-2}}$, $E\!M_{53} = 10^{53}$~cm$^{-3}$, and $F_{-13}=
  10^{-13}$~\ecms.The spectral fits were carried out in the
  energy range 0.3--7.5~keV.

    \leavevmode
    \begin{tabular}{r|cccccccccc}
Obs. & $N({\rm H})$ & $kT_1$ & $Z$ & $E\!M_1$ & $kT_2$ & $E\!M_2$ & $\chi^2$ & $P$ & $F_{\rm X}$ & $F_{\rm X}$ Intr.\\
\hline
~ &  $N_{22}$ & keV &  $Z_{\odot}$ & $E\!M_{53}$ & keV &  $E\!M_{53}$ & ~ & ~ & $F_{-13}$  & $F_{-13}$ \\
\hline

0201 & 0.13 $\pm$ 0.03 & 0.26 $\pm$ 0.04 & 0.05 $\pm$ 0.02     & 12.89 $\pm$ 11.85 & 1.19 $\pm$ 0.12 & 17.47 $\pm$ 3.31 & 1.04 & 0.35 & 35.33 & 55.92\\
0301 & 0.09 $\pm$ 0.02 & 0.32 $\pm$ 0.02 & 0.22 $\pm$ 0.06      & 4.67$\pm$ 2.61 & 1.48 $\pm$ 0.09  & 9.69 $\pm$ 1.40  & 1.27 & 0.01 & 43.09 & 45.96\\
0401 & 0.11 $\pm$ 0.02 & 0.30 $\pm$ 0.03 & 0.10 $\pm$ 0.04     & 5.55 $\pm$ 4.27 & 1.18 $\pm$ 0.09 & 7.92 $\pm$ 1.80  & 1.01 & 0.45 & 19.98 & 30.02\\
0501 & 0.06 $\pm$ 0.03 & 0.39 $\pm$ 0.07 & 0.20 $\pm$ 0.08     & 2.16 $\pm$ 1.94  & 1.30 $\pm$ 0.07  & 5.10 $\pm$ 1.62   & 0.91 & 0.78 & 18.54 & 22.80\\
0601 & 0.08 $\pm$ 0.02 & 0.42 $\pm$ 0.12 & 0.10 $\pm$ 0.03  & 2.07 $\pm$ 2.01 & 1.09 $\pm$ 0.07  & 5.85 $\pm$ 4.85  & 0.80 & 0.90 & 15.00 & 20.17\\
0701 & 0.06 $\pm$ 0.05 & 0.43 $\pm$ 0.14 & 0.23 $\pm$ 0.12     & 1.70 $\pm$ 2.37  & 1.28 $\pm$ 0.11  & 3.51 $\pm$ 1.71  & 1.00 & 0.45 & 13.91 & 17.43\\
0801 & 0.09 $\pm$ 0.04 & 0.36 $\pm$ 0.07 & 0.17 $\pm$ 0.09     & 2.82 $\pm$ 3.32  & 1.32 $\pm$ 0.10 & 4.05 $\pm$ 1.49   & 1.10 & 0.31 & 13.80 & 19.02\\
0901 & 0.15 $\pm$ 0.05 & 0.27 $\pm$ 0.04 & 0.12 $\pm$ 0.06     & 6.21 $\pm$ 7.82  & 1.09 $\pm$ 0.15  & 6.30 $\pm$ 2.84   & 0.99 & 0.51 & 14.91 & 26.25\\
1001 & 0.04 $\pm$ 4.90E-03 & 0.79 $\pm$ 0.06 & 0.93 $\pm$ 0.14 & 0.78 $\pm$ 0.31  & 5.08 $\pm$ 0.24   & 29.86 $\pm$ 1.11 & 0.99 & 0.57 & 200.39 & 219.25 \\
1101 & 0.05 $\pm$ 6.33E-03 & 0.76 $\pm$ 0.02 & 0.50 $\pm$ 0.09 & 2.42 $\pm$ 0.81 & 3.04 $\pm$ 0.14   & 13.65 $\pm$ 0.81  & 1.03 & 0.37 & 80.36 & 89.18\\
1201 & 0.11 $\pm$ 0.02 & 0.33 $\pm$ 0.03 & 0.14 $\pm$ 0.06      & 5.14 $\pm$ 4.03  & 1.45 $\pm$ 0.13  & 8.48 $\pm$ 1.68  & 1.19 & 0.08 & 25.53 & 36.67\\

    \end{tabular}
    \label{tab:v827_all11pn}
  \end{center}
\end{table*}

\begin{table*}[thbp]
  \begin{center}

  \caption{Best-fit spectral parameters of the PN data for the eleven
  exposures of V1075~Tau. In this case $Z$ was frozen at
  0.15~$Z_{\odot}$, the value derived for this source from the \xmm\
  observation of L1551 of 2000 (FGM03). $E\!M$ is the emission
  measure, $P$ the null-hypothesis probability of the fit, and $F_{\rm
  X}$ Intr. the "unabsorbed" X-ray flux. Units are $N_{22} =
  10^{22}~{\rm cm^{-2}}$, $E\!M_{53} = 10^{53}$~cm$^{-3}$, and
  $F_{-13}= 10^{-13}$~\ecms.The spectral fits were carried out in the
  energy range 0.3--7.5~keV.}

    \leavevmode
    \begin{tabular}{r|ccccccc}
Obs. & $N({\rm H})$ & $kT$ & $E\!M$ & $\chi^2$ & $P$ & $F_{\rm X}$ & $F_{\rm X}$ Intr.\\
\hline
~ &  $N_{22}$ & keV &  $E\!M_{53}$ & ~ & ~ & $F_{-13}$  & $F_{-13}$ \\
\hline

0201 & 0.06 $\pm$ 0.03 & 0.77 $\pm$ 0.04 & 2.04 $\pm$ 0.50 & 0.81 & 0.86 & 4.90 & 6.37\\
0301 & 0.07 $\pm$ 0.01 & 1.05 $\pm$ 0.02 & 5.30 $\pm$ 0.38 & 1.62 & 1.8E-4 & 13.06 & 17.48\\
0401 & 0.04 $\pm$ 0.02 & 0.95 $\pm$ 0.04 & 2.64 $\pm$ 0.35 & 0.94 & 0.6 & 7.32 & 8.61\\
0501 & 0.07 $\pm$ 0.03 & 0.88 $\pm$ 0.07 & 2.18 $\pm$ 0.54 & 1.09 & 0.25 & 5.15 & 7.05\\
0601 & 0.06 $\pm$ 0.02 & 0.94 $\pm$ 0.04 & 2.43 $\pm$ 0.35 & 0.84 & 0.76 & 6.16 & 7.91\\
0701 & 0.03 $\pm$ 0.03 & 1.03 $\pm$ 0.05 & 3.26 $\pm$ 0.59 & 1.51 & 0.09 & 9.17 & 10.15\\
0801 & 0.01 $\pm$ 0.03 & 0.76 $\pm$ 0.04 & 2.11 $\pm$ 0.54 & 0.89 & 0.61 & 6.24 & 6.56\\
0901 & 0.03 $\pm$ 0.03 & 0.76 $\pm$ 0.04 & 1.96 $\pm$ 0.44 & 0.99 & 0.53 & 5.19 & 6.13\\
1001 & 0.08 $\pm$ 0.03 & 0.75 $\pm$ 0.04 & 2.52 $\pm$ 0.57 & 1.41 & 0.02 & 5.41 & 7.81\\
1101 & 0.05 $\pm$ 0.01 & 1.05 $\pm$ 0.03 & 2.35 $\pm$ 0.26 & 1.96 & 9.2E-5 & 6.36 & 7.77\\
1201 & 0.11 $\pm$ 0.04 & 0.64 $\pm$ 0.04 & 2.33 $\pm$ 0.67 & 1.35 & 0.08 & 4.20 & 6.83\\

    \end{tabular}
    \label{tab:v1075_all11pn}
  \end{center}
\end{table*}

\begin{table*}[thbp]
  \begin{center}

  \caption{Spectral parameters derived from simultaneously fitting PN,
  MOS1, and MOS2 data for the eleven exposures of V710 Tau. In this
  case $Z$ was frozen at 0.2~$Z_{\odot}$. $E\!M$ is the emission
  measure, $P$ the null-hypothesis probability of the fit, and $F_{\rm
  X}$ Intr. the "unabsorbed" X-ray flux. Units are $N_{22} =
  10^{22}~{\rm cm^{-2}}$, $E\!M_{53} = 10^{53}$~cm$^{-3}$, and
  $F_{-13}= 10^{-13}$~\ecms. The spectral fits were carried out in
  the energy range 0.3--7.5~keV.}

    \leavevmode
    \begin{tabular}{r|ccccccc}
Obs. & $N({\rm H})$ & $kT$ & $E\!M$ & $\chi^2$ & $P$ & $F_{\rm X}$ & $F_{\rm X}$ Intr.\\
\hline
~ &  $N_{22}$ & keV &  $E\!M_{53}$ & ~ & ~ & $F_{-13}$  & $F_{-13}$ \\
\hline

0201 & 0.24 $\pm$ 0.23 & 0.90 $\pm$ 0.24 & 0.56 $\pm$ 0.52 & 0.9 & 0.63 & 0.98 & 1.86\\
0301 & 0.43 $\pm$ 0.10 & 0.83 $\pm$ 0.08 & 1.15 $\pm$ 0.58 & 1.35 & 0.14 & 1.31 & 3.71\\
0401 & 0.22 $\pm$ 0.13 & 0.78 $\pm$ 0.10 & 0.54 $\pm$ 0.40 & 0.71 & 0.76 & 0.91 & 1.74\\
0501 & 0.71 $\pm$ 0.27 & 0.41 $\pm$ 0.24 & 4.24 $\pm$ 12.66 & 1.35 & 0.05 & 1.01 & 9.93\\
0601 & 0.24 $\pm$ 0.12 & 0.78 $\pm$ 0.09 & 0.76 $\pm$ 0.49 & 1.12 & 0.34 & 1.23 & 2.39\\
0701 & 0.31 $\pm$ 0.37 & 0.60 $\pm$ 0.73 & 0.82 $\pm$ 3.73 & 0.71 & 0.54 & 0.86 & 2.03\\
0801 & 0.46 $\pm$ 0.21 & 0.72 $\pm$ 0.16 & 1.12 $\pm$ 1.26 & 0.95 & 0.48 & 1.07 & 3.16\\
0901 & 0.10 $\pm$ 0.18 & 0.68 $\pm$ 0.15 & 0.35 $\pm$ 0.48 & 1.23 & 0.10 & 0.81 & 1.25\\
1001 & 0.33 $\pm$ 0.17 & 0.75 $\pm$ 0.11 & 0.86 $\pm$ 0.76 & 1.22 & 0.21 & 1.11 & 2.69\\
1101 & 0.17 $\pm$ 0.07 & 1.03 $\pm$ 0.06 & 0.82 $\pm$ 0.25 & 1.61 & 0.04 & 1.73 & 2.67\\
1201 & 0.40 $\pm$ 0.12 & 0.68 $\pm$ 0.09 & 1.29 $\pm$ 0.90 & 1.18 & 0.27 & 1.32 & 3.79\\

    \end{tabular}
    \label{tab:v710_all11pnmos12}
  \end{center}
\end{table*}

\begin{table*}[thbp]
  \begin{center}

  \caption{Best-fit spectral parameters of the PN data for the nine
  exposures of XZ Tau. In this case $Z$ was frozen at
  0.08~$Z_{\odot}$. $E\!M$ is the emission measure, $P$ the
  null-hypothesis probability of the fit, and $F_{\rm X}$ Intr. the
  "unabsorbed" X-ray flux. Units are $N_{22} = 10^{22}~{\rm
  cm^{-2}}$, $E\!M_{53} = 10^{53}$~cm$^{-3}$, and $F_{-13}=
  10^{-13}$~\ecms. The spectral fits were carried out in the
  energy range 0.3--7.5~keV. The quality of data was such that the
  parameter space of the model was degenerate in $N({\rm H})$ and $Z$,
  and changing the initial values could result in significantly
  different best-fit values for these two parameters. The value
  $Z=0.08~Z_{\odot}$ was determined by first performing a fit in which
  $Z$ was allowed to vary, determining the average value over the 11
  exposures and then repeating the spectral fits with $Z$ frozen. }

    \leavevmode
    \begin{tabular}{r|ccccccc}
Obs. & $N({\rm H})$ & $kT$ & $E\!M_1$ & $\chi^2$ & $P$ & $F_{\rm X}$ & $F_{\rm X}$ Intr.\\
\hline
~ &  $N_{22}$ & keV &  $E\!M_{53}$ & ~ & ~ & $F_{-13}$  & $F_{-13}$ \\
\hline
0201 & 0.29 $\pm$ 0.06 & 0.63 $\pm$ 0.06 & 3.02 $\pm$ 1.31 & 1.07 & 0.36 & 2.27 & 6.69\\
0301 & 0.15 $\pm$ 0.03 & 0.78 $\pm$ 0.04 & 1.46 $\pm$ 0.32 & 0.78 & 0.78 & 1.98 & 3.65\\
0401 & 0.08 $\pm$ 0.02 & 0.77 $\pm$ 0.04 & 1.25 $\pm$ 0.26 & 1.38 & 0.08 & 2.15 & 3.11\\
0501 & 0.11 $\pm$ 0.04 & 0.72 $\pm$ 0.05 & 1.47 $\pm$ 0.41 & 1.17 & 0.20 & 2.13 & 3.53\\
0601 & 0.07 $\pm$ 0.02 & 0.78 $\pm$ 0.04 & 1.19 $\pm$ 0.24 & 1.02 & 0.43 & 2.13 & 2.97\\
0901 & 0.14 $\pm$ 0.03 & 0.76 $\pm$ 0.05 & 1.68 $\pm$ 0.44 & 1.09 & 0.24 & 2.28 & 4.15\\
1001 & 0.09 $\pm$ 0.02 & 1.03 $\pm$ 0.07 & 1.45 $\pm$ 0.26 & 1.15 & 0.23 & 2.83 & 4.06\\
1101 & 0.14 $\pm$ 0.02 & 1.00 $\pm$ 0.04 & 2.37 $\pm$ 0.25 & 1.56 & 4.87E-03 & 3.89 & 6.56\\
1201 & 0.13 $\pm$ 0.02 & 0.80 $\pm$ 0.04 & 1.80 $\pm$ 0.33 & 1.15 & 0.24 & 2.64 & 4.55\\
    \end{tabular}
    \label{tab:xz_all11pn}
  \end{center}
\end{table*}

\begin{table*}[thbp]
  \begin{center}

  \caption{Best-fit spectral parameters of the PN data for the eleven
  exposures of HL Tau. In this case $Z$ was frozen at 0.6~$Z_{\odot}$,
  the value derived for this source from the \xmm\ observation of
  L1551 of 2000 (FGM03). Spectral fits were limited to energy range
  1.0--7.5~keV. $E\!M$ is the emission measure, $P$ the
  null-hypothesis probability of the fit, and $F_{\rm X}$ Intr. the
  "unabsorbed" X-ray flux. Units are $N_{22} = 10^{22}~{\rm
  cm^{-2}}$, $E\!M_{53} = 10^{53}$~cm$^{-3}$, and $F_{-13}=
  10^{-13}$~\ecms. }

    \leavevmode
    \begin{tabular}{r|ccccccc}
Obs. & $N({\rm H})$ & $kT$ & $E\!M$ & $\chi^2$ & $P$ & $F_{\rm X}$ & $F_{\rm X}$ Intr.\\
\hline
~ &  $N_{22}$ & keV &  $E\!M_{53}$ & ~ & ~ & $F_{-13}$  & $F_{-13}$ \\
\hline

0201 & 3.52 $\pm$ 1.17 & 2.73 $\pm$ 1.16 & 2.36 $\pm$ 2.29 & 0.5 & 0.89 & 3.8 & 7.8\\
0301 & 2.19 $\pm$ 0.58 & 4 $\pm$ 1.78 & 0.79 $\pm$ 0.56 & 0.89 & 0.51 & 2.1 & 3.7\\
0401 & 1.47 $\pm$ 0.8 & 5.41 $\pm$ 3.33 & 0.96 $\pm$ 0.76 & 1.45 & 0.14 & 3.4 & 4.4\\
0501 & 3.98 $\pm$ 1.94 & 2.48 $\pm$ 1.46 & 1.59 $\pm$ 2.57 & 0.64 & 0.84 & 2.2 & 6.0\\
0601 & 3.04 $\pm$ 0.44 & 1.75 $\pm$ 0.27 & 2.66 $\pm$ 1.23 & 0.7 & 0.73 & 2.7 & 8.9\\
0701 & 3.14 $\pm$ 0.19 & 9.1 $\pm$ 1.83 & 14.08 $\pm$ 1.03 & 0.75 & 0.9 & 48.4 & 81.9\\
0801 & 3.14 $\pm$ 0.23 & 4.56 $\pm$ 0.67 & 12.75 $\pm$ 1.75 & 1.03 & 0.42 & 33.2 & 61.5\\
0901 & 3.41 $\pm$ 0.2 & 8.53 $\pm$ 0.96 & 8.46 $\pm$ 0.70 & 0.93 & 0.71 & 28.0 & 48.9\\
1001 & 2.87 $\pm$ 0.53 & 2.41 $\pm$ 0.53 & 3.41 $\pm$ 1.73 & 0.59 & 0.95 & 5.2 & 12.7\\
1101 & 2.12 $\pm$ 0.39 & 4.11 $\pm$ 1.22 & 1.34 $\pm$ 0.63 & 1.49 & 0.11 & 3.7 & 6.3\\
1201 & 2.08 $\pm$ 0.49 & 3.4 $\pm$ 1.16 & 1.27 $\pm$ 0.79 & 1.65 & 0.09 & 3.1 & 5.6\\
    \end{tabular}
    \label{tab:hl_all11pn}
  \end{center}
\end{table*}

\begin{table*}[thbp]
  \begin{center} \caption{Best-fit spectral parameters of the EPIC-PN
  data for the eleven exposures of HD285845. $E\!M$ is the
  emission measure, $P$ the null-hypothesis probability of the fit, and
  $F_{\rm X}$ Intr. the "unabsorbed" X-ray flux. Units are $N_{22} =
  10^{22}~{\rm cm^{-2}}$, $E\!M_{53} = 10^{53}$~cm$^{-3}$, and
  $F_{-13}= 10^{-13}$~\ecms. The metallicity was frozen at
  $Z=0.14~Z_{\sun}$. The spectral fits were carried out in the
  energy range 0.3--7.5~keV.}
    \leavevmode
        \footnotesize
    \begin{tabular}{r|ccccccccc}
Obs. & $N({\rm H})$ & $kT_1$ & $E\!M_1$ & $kT_2$ & $E\!M_2$ & $\chi^2$ & $P$ & $F_{\rm X}$ & $F_{\rm X}$ Intr.\\
\hline
~ &  $N_{22}$ & keV & $E\!M_{53}$ & keV &  $E\!M_{53}$ & ~ & ~ & $F_{-13}$  & $F_{-13}$ \\
\hline
0201 & 0.07 $\pm$ 0.02 & 0.41 $\pm$ 0.06 & 2.51 $\pm$ 1.34 & 0.97 $\pm$ 0.03 & 4.94 $\pm$ 0.55 & 1.07 & 0.29 & 15.16 & 20.99\\
0301 & 0.11 $\pm$ 0.02 & 0.34 $\pm$ 0.02 & 4.79 $\pm$ 1.39 & 0.96 $\pm$ 0.03 & 3.98 $\pm$ 0.39 & 1.07 & 0.27 & 12.77 & 21.19\\
0401 & 0.09 $\pm$ 0.01 & 0.34 $\pm$ 0.02 & 4.71 $\pm$ 1.34 & 1.00 $\pm$ 0.03 & 5.19 $\pm$ 0.42 & 1.04 & 0.33 & 16.13 & 25.02\\
0501 & 0.11 $\pm$ 0.02 & 0.36 $\pm$ 0.04 & 4.29 $\pm$ 1.84 & 1.02 $\pm$ 0.04 & 3.64 $\pm$ 0.54 & 0.98 & 0.57 & 12.08 & 19.77\\
0601 & 0.07 $\pm$ 0.01 & 0.34 $\pm$ 0.03 & 3.45 $\pm$ 1.16 & 0.98 $\pm$ 0.02 & 5.57 $\pm$ 0.41 & 0.99 & 0.52 & 16.87 & 23.96\\
0701 & 0.07 $\pm$ 0.02 & 0.35 $\pm$ 0.05 & 4.04 $\pm$ 1.96 & 1.19 $\pm$ 0.05 & 6.26 $\pm$ 0.61 & 1.08 & 0.3 & 20.71 & 28.09\\
0801 & 0.09 $\pm$ 0.02 & 0.35 $\pm$ 0.03 & 5.09 $\pm$ 2.05 & 1.02 $\pm$ 0.06 & 3.40 $\pm$ 0.59 & 0.87 & 0.76 & 12.73 & 20.09\\
0901 & 0.08 $\pm$ 0.02 & 0.34 $\pm$ 0.03 & 4.01 $\pm$ 1.39 & 1.01 $\pm$ 0.03 & 4.86 $\pm$ 0.47 & 1.14 & 0.03 & 15.46 & 22.71\\
1001 & 0.07 $\pm$ 0.01 & 0.39 $\pm$ 0.03 & 4.11 $\pm$ 1.39 & 1.19 $\pm$ 0.04 & 5.42 $\pm$ 0.44 & 1.27 & 0.01 & 19.03 & 26.05\\
1101 & 0.08 $\pm$ 0.01 & 0.32 $\pm$ 0.02 & 3.81 $\pm$ 1.05 & 1.05 $\pm$ 0.03 & 4.35 $\pm$ 0.33 & 0.85 & 0.93 & 13.81 & 20.58\\
1201 & 0.08 $\pm$ 0.01 & 0.42 $\pm$ 0.06 & 2.33 $\pm$ 1.13 & 0.97 $\pm$ 0.03 & 5.33 $\pm$ 0.45 & 0.99 & 0.52 & 15.56 & 21.99\\
    \end{tabular}
    \label{tab:hd_all11pn}
  \end{center}
\end{table*}

\end{document}